%% file: alicepreprint_ewBoson_pPb.tex
\documentclass[ALICE,manyauthors]{cernphprep}

\usepackage[comma,square,numbers,sort&compress]{natbib}
\usepackage{hyperref}
\usepackage{lineno}

\input{texCommands}

\begin{document}%

\begin{titlepage}
\PHyear{2016}
\PHnumber{278}      
\PHdate{4 November}  
%

\title{W and Z boson production in \pPb collisions at \texorpdfstring{\snn}{sqrt(s\_NN)}=5.02~TeV}

\Collaboration{ALICE Collaboration\thanks{See Appendix~\ref{app:collab} for the list of collaboration members}}
\ShortAuthor{ALICE Collaboration} 

\begin{abstract}
\input{abstract_ewBoson_pPb}
\end{abstract}
\end{titlepage}
\setcounter{page}{2}

\input{ewBoson_pPb}               

\newenvironment{acknowledgement}{\relax}{\relax}
\begin{acknowledgement}
\section*{Acknowledgements}
\input{fa_2016-11-04.tex}    
\end{acknowledgement}

\bibliographystyle{utphys}   
\bibliography{biblio}

\newpage
\appendix
\section{The ALICE Collaboration}
\label{app:collab}
\input{Alice_Authorlist_2016-09-23_mod.tex}  
\end{document}

%% file: texCommands.tex
\newcommand{\snn}{\ensuremath{\sqrt{s_{\mathrm{NN}}}}\xspace}

\newcommand{\eg}{e.g.\,}
\newcommand{\wBoson}[1][\pm]{\ensuremath{\text{W}^{#1}}\xspace}

\newcommand{\ncoll}{\ensuremath{N_\text{coll}}\xspace}
\newcommand{\ncmult}{\ensuremath{N_\text{coll}^\text{mult}}\xspace}
\newcommand{\npart}{\ensuremath{N_\text{part}}\xspace}
\newcommand{\npmult}{\ensuremath{N_\text{part}^\text{mult}}\xspace}

\newcommand{\sect}[1]{Section~\ref{#1}\xspace}
\newcommand{\pt}{\ensuremath{p_{\textrm{T}}}\xspace}

\newcommand{\ycms}[1][]{\ensuremath{y_{\text{cms}}^{#1}}\xspace}
\newcommand{\lumi}[1][n]{~\ensuremath{\text{#1b}^{-1}}\xspace}
\newcommand{\GeVc}{~\mbox{GeV/}\ensuremath{c}\xspace}
\newcommand{\GeVcc}{~\mbox{GeV/}\ensuremath{c^2}\xspace}

\newcommand{\nsub}[1]{\ensuremath{N_{\rm #1}}\xspace}
\newcommand{\fsub}[1]{\ensuremath{f_{\rm #1}}\xspace}
\newcommand{\wmu}{\ensuremath{\mu \leftarrow \rm W}\xspace}
\newcommand{\wmus}[1][\pm]{\ensuremath{\mu^#1 \leftarrow \mathrm{W}^#1}}
\newcommand{\zmu}{\ensuremath{\mu \leftarrow \mathrm{Z}}\xspace}
\newcommand{\fitNbkg}{\ensuremath{N_\text{bkg}^\text{raw}}\xspace}
\newcommand{\fitNw}{\ensuremath{N_\text{\wmu}^\text{raw}}\xspace}
\newcommand{\eq}[1]{Eq.~\ref{#1}\xspace}
\newcommand{\fig}[1]{\figurename~\ref{#1}\xspace}
\newcommand{\tab}[1]{\tablename~\ref{#1}\xspace}

\newcommand{\acceff}{\ensuremath{A \times \epsilon}\xspace}

\newcommand{\invmass}{\ensuremath{m_{\mu\mu}}\xspace}
\newcommand{\pPb}{p--Pb\xspace}
\newcommand{\PbPb}{Pb--Pb\xspace}
\newcommand{\fnorm}[1][]{\ensuremath{F_{#1\mu\text{-trig/MB}}}\xspace}

\newcommand{\mum}{~\ensuremath{\mu\text{m}}\xspace}

\graphicspath{{Plots/}}

%% file: abstract_ewBoson_pPb.tex
The W and Z boson production was measured via the muonic decay channel in proton--lead collisions at \snn = 5.02 TeV at the Large Hadron Collider with the ALICE detector.
The measurement covers backward ($-4.46 < \ycms < -2.96$) and forward ($2.03 < \ycms < 3.53$) rapidity regions, corresponding to Pb-going and p-going directions, respectively.
The Z-boson production cross section, with dimuon invariant mass of $60<\invmass<120$\GeVcc and muon transverse momentum ($\pt^\mu$) larger than 20\GeVc, is measured.
The production cross section and charge asymmetry of muons from W-boson decays with $\pt^\mu>10$\GeVc are determined.
The results are compared to theoretical calculations both with and without including the nuclear modification of the parton distribution functions.
The W-boson production is also studied as a function of the collision centrality: the cross section of muons from W-boson decays is found to scale with the average number of binary nucleon-nucleon collisions within uncertainties.

%% file: ewBoson_pPb.tex
\section{Introduction}
The W and Z boson production is extensively studied at hadron colliders as it represents an important benchmark of the Standard Model.
The measurements in pp and p$\overline{\rm p}$ collisions at different energies~\cite{Albajar:1987yz,Alitti:1991dm,Abe:1995bm,Abulencia:2005ix,Abbott:1999tt,Adare:2010xa,Aggarwal:2010vc,Aad:2011dm,Aad:2016naf,CMS:2011aa,Chatrchyan:2014mua,Aaij:2015gna,Aaij:2015zlq} are well described by Quantum Chromodynamics (QCD) calculations at Next-to-Leading Order (NLO) and Next-to-Next-to-Leading Order (NNLO) in perturbation theory.
In the calculations, the input electroweak parameters (\eg boson masses and weak couplings) are known to high accuracy, as well as the radiative corrections~\cite{Baur:1998kt}.
The measurements can hence constrain the Parton Distribution Functions (PDFs)~\cite{Martin:1999ww}.

With the large centre-of-mass energies and luminosity of the Large Hadron Collider (LHC), the W and Z boson production has become accessible for the first time in proton-nucleus~\cite{Aad:2015gta,Khachatryan:2015pzs,Aaij:2014pvu,Khachatryan:2015hha} and nucleus-nucleus collisions~\cite{Aad:2014bha,Aad:2012ew,Chatrchyan:2012nt,Chatrchyan:2014csa}.
The PDFs are expected to be modified for nucleons inside a nucleus compared to those of nucleons in vacuum.
Nuclear PDFs (nPDFs) are extracted from global analyses performed at NLO accuracy in perturbative QCD~\cite{deFlorian:2011fp,Eskola:2009uj},
but the results are mostly constrained by Deep-Inelastic Scattering and Drell-Yan data in a limited region of the four-momentum transfer $Q^2$ and parton longitudinal momentum fraction Bjorken-$x$~\cite{Eskola:2009uj}.
The W and Z bosons and their lepton decay products are unaffected by the hot and dense strongly-interacting matter formed in ultra-relativistic heavy-ion collisions and offer a unique opportunity to study the nPDF in a region of high $Q^2 \sim (100~\text{GeV})^2$ and Bjorken-$x$ ranges from $\sim 10^{-4}$ to almost unity where they are poorly constrained by data~\cite{Paukkunen:2010qg}.
Furthermore, the asymmetry in the production of positive and negative W bosons, occurring mainly in the processes $\mathrm{u\overline{d}} \rightarrow \wBoson[+]$ and $\mathrm{d\overline{u}} \rightarrow \wBoson[-]$ at the LHC energies, can be used to probe the flavour modification of the quark densities in nuclei~\cite{Paukkunen:2010qg}.

The W and Z boson production was measured in \PbPb collisions at \snn = 2.76~TeV by the ATLAS~\cite{Aad:2014bha,Aad:2012ew} and the CMS~\cite{Chatrchyan:2012nt,Chatrchyan:2014csa} experiments in the electronic and muonic decay channels.
The results confirm that the production cross section scales with the number of nucleon-nucleon collisions (binary scaling) within uncertainties on the order of 10\%.
The W and Z bosons were further studied in \pPb collisions at \snn = 5.02~TeV.
The Z-boson production was measured by the ATLAS~\cite{Aad:2015gta} and CMS~\cite{Khachatryan:2015pzs} experiments at mid-rapidity in the leptonic decay channels, and by the LHCb experiment at forward rapidities~\cite{Aaij:2014pvu} in the muonic decay channel.
The W-boson production was measured by the CMS experiment at mid-rapidity~\cite{Khachatryan:2015hha} in the leptonic (e, $\mu$) decay channel.
The results are described by theoretical calculations both with and without including the nuclear modification of the PDFs, with a preference towards the former and can be used to further constrain the nPDFs~\cite{Paukkunen:2010qg}.

In nucleus-nucleus collisions, particle production is often studied as a function of the collision centrality, which is directly related to the impact parameter of the collision.
The number of interacting nucleons, and hence the energy deposited in the collision region, increases from peripheral to central (head-on) collisions thus affecting the volume and density of the strongly-interacting medium that is produced.
The nuclear modification of the PDFs is expected to depend as well on the position of the nucleon inside the nucleus, and therefore on average on the impact parameter of the collision~\cite{Helenius:2012wd}.
The centrality of nucleus-nucleus collisions is usually estimated by measuring either the energy deposition or the hadronic multiplicity in specific detectors.
This estimation is known to be biased in \pPb collisions, where the range of the multiplicity is of similar magnitude as its fluctuations~\cite{Adam:2014qja}.
The biases are minimised when the centrality is determined through the energy measured at beam rapidity (with zero degree calorimeters), which is deposited by the non-interacting (spectator) nucleons emitted from the Pb nucleus in the collision and is therefore independent of the fluctuations in the number of produced particles.

The W and Z boson production occurs in hard scattering processes at the initial stage of the collision, and it is expected to scale with the number of binary nucleon-nucleon collisions.
The centrality-dependent yield can be therefore used as a test bench for the centrality estimation at the LHC.

In this article, the ALICE results on Z and W boson production in the muonic decay channel in  \pPb collisions at \snn = 5.02~TeV are presented.
The former is measured with smaller uncertainties than the corresponding LHCb measurement in a similar rapidity range.
The latter is the first measurement of W production in \pPb collisions at forward and backward rapidity, in a region that is complementary to the one explored by CMS.
The article is organized as follows. The data sample and analysis strategies are described in \sect{sec:analysis}.
The results are shown in \sect{sec:results} and summarised in \sect{sec:conclusions}.

\section{Data analysis}\label{sec:analysis}
\subsection{Experimental apparatus and data samples}\label{sec:dataSample}
The ALICE detector is described in detail in~\cite{Aamodt:2008zz}.
Muons are reconstructed in the muon spectrometer, covering the pseudorapidity range $-4<\eta<-2.5$ in the laboratory frame.
The spectrometer consists of a dipole magnet with a 3 Tm integrated magnetic field, five tracking stations made of Multi-Wire Proportional Chambers with Cathode Pad readout, and two trigger stations made of Resistive Plate Chambers and several absorption elements.
The tracking stations are placed downstream from a conical front absorber made of carbon, concrete and steel, with a thickness of 4.1~m (corresponding to 10 nuclear interaction lengths, $\lambda_\text{I}$) that filters out hadrons from the interaction point.
The trigger stations are placed after an iron wall with a thickness of 1.2~m (7.2~$\lambda_\text{I}$) that absorbs secondary hadrons escaping from the front absorber and low-momentum muons, mainly coming from the decay of light hadrons.
Finally, a conical beam shield covering the beam pipe protects the spectrometer from particles produced in the interaction of large-$\eta$ particles with the pipe itself.

In this analysis, the position of the interaction vertex is measured with the Silicon Pixel Detector (SPD), which constitutes the two innermost layers of the Inner Tracking System, covering an acceptance interval of $|\eta|<2$ and $|\eta|<1.4$, for the first and second layer, respectively.
Two arrays of scintillators, the V0 detector~\cite{Abbas:2013taa}, placed on each side of the interaction point and covering the pseudprapidity regions $2.8 < \eta < 5.1$ and $-3.7 < \eta < -1.7$, are used as trigger detectors and to reject beam-induced background.
The V0 is also used as a luminometer, together with the T0 detector, which consists of two arrays of quartz Cherenkov counters covering the pseudprapidity regions $4.6 < \eta < 4.9$ and
$-3.3 < \eta < -3.0$.
The neutron zero degree calorimeters (ZN), placed on either side of the interaction point at $\pm$112.5 m along the beam pipe are used to estimate the centrality of the collision.

The analysis is performed on data collected in 2013 in proton--lead collisions at a centre-of-mass energy $\snn = 5.02$~TeV.
Due to the different energies of the proton and lead beams ($E_\text{p} = 4$~TeV and $E_\text{Pb} = 1.58$~TeV per nucleon), the resulting nucleon--nucleon centre-of-mass is boosted with respect to the laboratory frame by $\Delta y = 0.465$ in the direction of the protons.
Data were collected in two configurations, by inverting the direction of the p and Pb beams.
It is assumed that the proton beam travels towards positive rapidities.
With this convention, muons are measured at forward rapidity ($2.03 < \ycms< 3.53$) when the proton travels towards the spectrometer and at backward rapidity ($-4.46 < \ycms < -2.96$) when the Pb ion is travelling towards the spectrometer.
In the following, the two configurations will be referred to as p-going and Pb-going directions, respectively.

The data sample used in the W-boson analysis consists of events with at least one muon candidate track selected with the muon trigger with a transverse momentum $\pt \gtrsim4.2$\GeVc, in coincidence with a Minimum Bias (MB) event, which is defined by requiring the coincidence of signals in the two arrays of the V0 detector.
For the Z-boson analysis, two muon candidates with a transverse momentum of $\pt \gtrsim 0.5$\GeVc are required, in coincidence with a MB event.
The trigger selection on the muon \pt is not sharp and the threshold is defined as the value for which the trigger efficiency reaches a value of 50\%.
The integrated luminosities used in the analysis were computed by estimating the equivalent number of MB events corresponding to the muon-triggered data samples and then dividing by the MB cross sections.
The latter were measured with Van der Meer scans and amount to $2.12 \pm 0.07$~b and $2.09 \pm 0.07$~b for the Pb-going and p-going samples, respectively~\cite{Abelev:2014epa}.
The number of MB events corresponding to the muon-triggered data sample is evaluated as $N_\text{MB} = \fnorm \cdot N_{\mu-\text{trig}}$ where $N_{\mu-\text{trig}}$ is the number of muon-triggered events and \fnorm is the inverse probability of having a muon-triggered event in a MB event.
The normalisation factor \fnorm is estimated by using the information of the counters recording the total number of triggers, corrected for pile-up effects, which amount to 2\%.
The \fnorm factor can also be obtained by applying the muon trigger condition in the analysis of MB events.
The difference between the results obtained with the two methods, which amounts to about 1\%, is taken as the systematic uncertainty.
The integrated luminosity was also independently measured using the T0 detector: the results agree within better than 1\% in both data samples.
The difference was included in the systematic uncertainty of the MB cross section.
The resulting luminosity is $5.81 \pm 0.20$\lumi and $5.03 \pm 0.18$\lumi for the Pb-going and p-going data samples, respectively.

The centrality of the collision is measured from the energy deposited in the ZN in the direction of the fragmenting lead ion.
The average number of binary nucleon-nucleon collisions $\langle \ncoll \rangle$ is obtained from the ``hybrid method'' described in~\cite{Adam:2014qja}, which relies on the assumption that the charged-particle multiplicity measured at mid-rapidity is proportional to the average number of nucleons participating in the interaction $\langle \npart \rangle$.
The values of $\langle \npart \rangle$ for a given ZN-centrality class are calculated by scaling the average number of participants in MB collisions $\langle \npart^\text{MB} \rangle$, estimated with a Glauber Monte Carlo~\cite{Miller:2007ri}, by the ratio of the average charged-particle multiplicity measured at mid-rapidity for the ZN-centrality class and that of MB.
These values are denoted as  $\langle \npmult \rangle$ in the following to indicate the assumption used for the scaling.
The corresponding number of binary collisions is then obtained as: $\langle \ncmult \rangle  = \langle \npmult \rangle - 1$.
The systematic uncertainties are estimated by using different ans\"atze, as described in~\cite{Adam:2014qja}.
The resulting values of $\langle \ncmult \rangle$ and their uncertainties are summarised in \tab{tab:ncoll}.

\begin{table}[t]
 \centering
 \begin{tabular}{|*{6}{c}|}
 \hline
 Centrality class & 0--100\% & 2--20\% & 20--40\% & 40--60\% & 60--100\% \\
 $\langle \ncmult \rangle$ & $6.9 \pm 0.6$ & $11.3 \pm 0.3$ & $9.6 \pm 0.2$ & $7.1 \pm 0.3$ & $3.2 \pm 0.1$\\
  \hline
 \end{tabular}
 \caption{Average number of binary nucleon-nucleon collisions $\langle \ncmult \rangle$ estimated with the hybrid ZN method~\cite{Adam:2014qja}.}
 \label{tab:ncoll}
\end{table}

The muon trigger efficiency is found to be independent of centrality in \pPb collisions.
The normalisation factor of muon-triggered to MB events per centrality class can be obtained from the centrality integrated value \fnorm scaled by the fraction of the MB events in the given centrality class.
The 0--2\% most central collisions are excluded in the centrality-dependent analysis, because of the large pile-up contamination in this event class (of the order of 20--30\%).
In pile-up events the ZN energies of two (or more) interactions sum up, thus biasing the centrality determination towards the most central classes.
The contamination is reduced with decreasing centrality, and is about 3\% in the 2--20\% event classes in both the p-going and Pb-going data sample.
These values are taken into account in the systematic uncertainties on the normalisation.

\subsection{Muon selection and Monte Carlo simulations}\label{sec:muonSelectionAndMC}

Muon track candidates are reconstructed in the tracking system using the algorithm described in~\cite{Aamodt:2011gj}.
A fiducial cut on the pseudorapidity of the muon of $-4<\eta<-2.5$ is applied in order to remove the particles at the edge of the spectrometer acceptance.
An additional selection on the polar angle measured at the end of the front absorber of $170^\circ<\theta_\text{abs}<178^\circ$ is required to reject muons crossing the high-density region of the front absorber that undergo significant scattering.
Muon identification is carried out by matching the tracks reconstructed in the tracker and the trigger systems.
The contamination from beam-induced background tracks, which do not point to the interaction vertex, can be efficiently removed by exploiting the correlation between the momentum ($p$) of the track and its Distance of Closest Approach (DCA) to the vertex.
Due to the multiple scattering in the front absorber, the DCA distribution of particles produced in the collision can be described with a Gaussian function, whose width depends on the material crossed  and is proportional to 1/$p$.
On the other hand, the background tracks have a DCA larger than about 40~cm, independent of \pt.
They can therefore be rejected by selecting particles with a $p\cdot$DCA smaller than 6 times the width of the distribution, extracted from a Gaussian fit.
The contamination depends on the beam configuration, being of the order of 7\% in the p-going direction and up to 90\% in the Pb-going direction for particles with $\pt>10\GeVc$.
However, in this region the signal and the background are completely separated and the selection can fully remove the background, with a signal rejection smaller than 0.3\%.

The probability of a cosmic muon to be reconstructed in coincidence with a minimum bias trigger is very small, of the order of $10^{-10}$.
The selection on the $p\cdot$DCA of the track further reduces the contamination to a negligible level.

The detector response for muons from W and Z boson decays was determined through Monte Carlo (MC) simulations.
The W and Z bosons are produced using POWHEG~\cite{Alioli:2008gx}, a NLO particle generator, paired with PYTHIA 6.425~\cite{Sjostrand:2006za} for parton shower.
The calculations include the CT10~\cite{Lai:2010vv} PDF set and the EPS09NLO~\cite{Eskola:2009uj} parameterisation of the nuclear modification of the PDFs.
The propagation of particles through the detector and the absorption materials uses the GEANT3~\cite{Brun:1994aa} transport code.
The simulation of \pPb collisions takes into account the isospin dependence (in terms of u- and d-type quark content) of the W and Z boson production, which is particularly important for W bosons~\cite{ConesaDelValle:2007dza}.
To this aim proton--proton (pp) and proton--neutron (pn) collisions are simulated separately.
The \pPb collisions are obtained as the sum of the results, weighted by the average number of pp and pn interactions in a \pPb collision.

The alignment of the tracking chambers is a crucial step in the analysis of muons at high transverse momentum.
The absolute position of the chambers was measured before data taking with photogrammetry.
Their relative position is estimated with a precision of about 100\mum, using a modified version of the MILLIPEDE~\cite{Blobel:2002ax} package, which combines data taken with and without the magnetic field.
The residual misalignment of the tracking chambers is taken into account in the simulations to estimate the acceptance and efficiency (\acceff) of the detector.
While the method provides the most accurate estimation of the relative chamber position, it is not sensitive to a global misalignment of the entire spectrometer.
A data-driven method was hence developed, in which the simulation of the tracker response is based on a parameterisation of the measured resolution of the clusters associated to a track.
The distribution of the difference between the cluster and the reconstructed track positions on each chamber is parameterised with an extended Crystal-Ball function~\cite{Gaiser:1982yw} and utilised to simulate the smearing of the track parameters.
The effect of a global misalignment of the muon spectrometer is mimicked by shifting the distribution of the track deviation in the magnetic field in opposite directions for positive and negative tracks.
This shift is tuned so as to reproduce the observed difference in the ratio of the \pt distributions of positive and negative tracks, corrected for acceptance and efficiency, in two periods of data taking differing only by the magnetic field polarity.
The values of the \acceff corrections are obtained using either the standard simulations with the residual misalignment, or the data-driven simulations: the difference is about 1\% (2\%) in the p-going (Pb-going) data sample for Z bosons, and about 1\% for W bosons.
These values are taken as the systematic uncertainties.
It is worth noting that the limited momentum resolution of the detector can also result in positive muons wrongly reconstructed as negative muons and viceversa.
The resulting loss of efficiency is small (smaller than 1\% for muons with $\pt>10$\GeVc) and taken into account in the simulations.

The uncertainty on the muon tracking efficiency is estimated from the difference between the muon tracking efficiency in MC and that from a data-driven approach based on the redundancy of the tracking stations~\cite{Abelev:2014ffa}.
It amounts to 2\% (3\%) for the p-going (Pb-going) period.
The uncertainty on trigger efficiency, which is mainly due to the systematic uncertainty in the determination of the efficiency of each trigger chamber from data, amounts to 1\%.
An additional systematic uncertainty of 0.5\% results from the choice of the $\chi^2$ cut in the matching of the tracks reconstructed in the tracker with those in the trigger.
In the dimuon analysis, these systematic uncertainties apply to both muons of the pair, which are well separated in phase space and therefore cross different parts of the detector.

\subsection{Z-boson analysis}\label{zAnalysis}

\begin{figure}[t]
\centering
\includegraphics[width=0.48\textwidth,trim=20 5 45 30,clip]{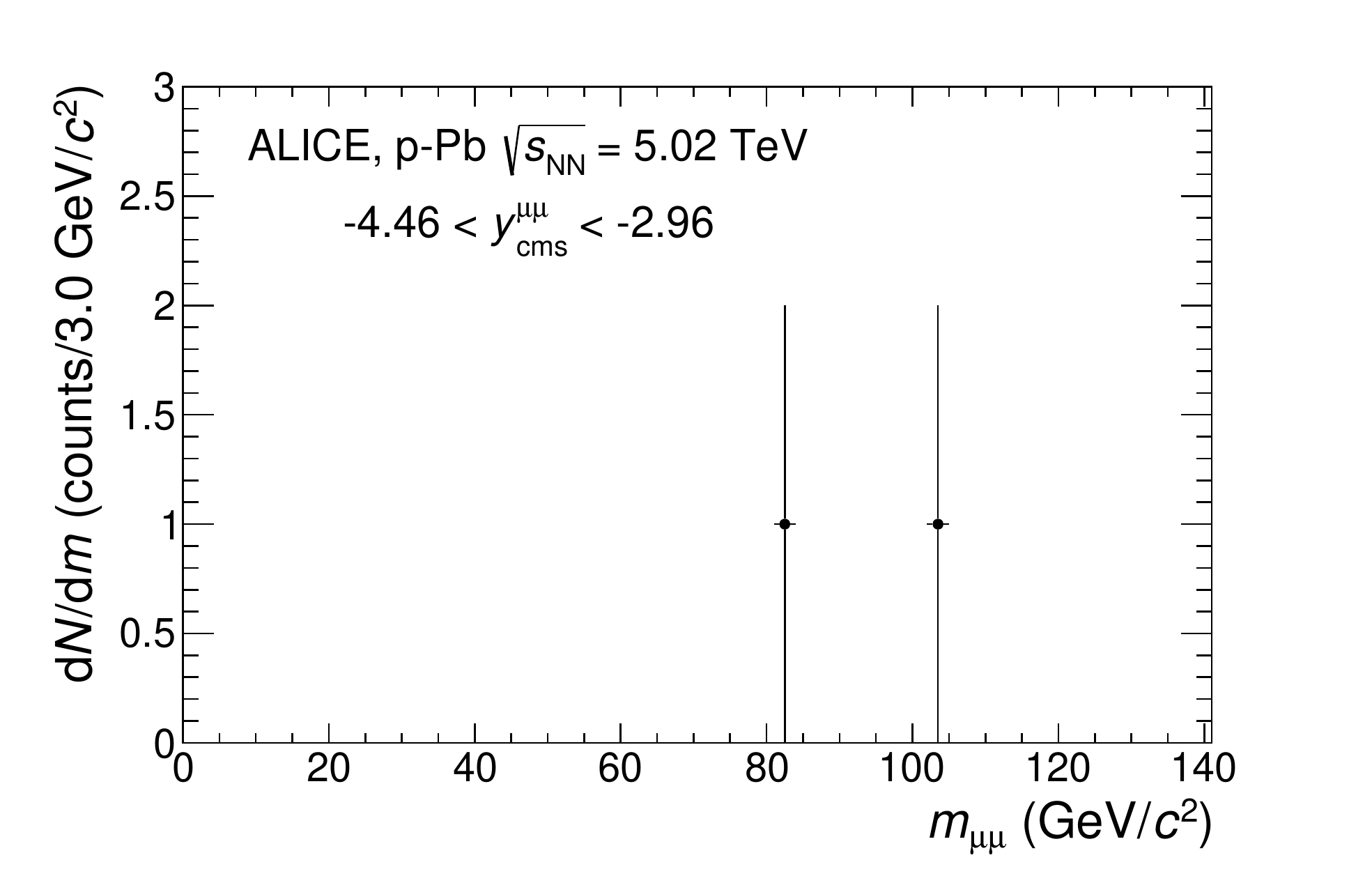}
\includegraphics[width=0.48\textwidth,trim=20 5 45 30,clip]{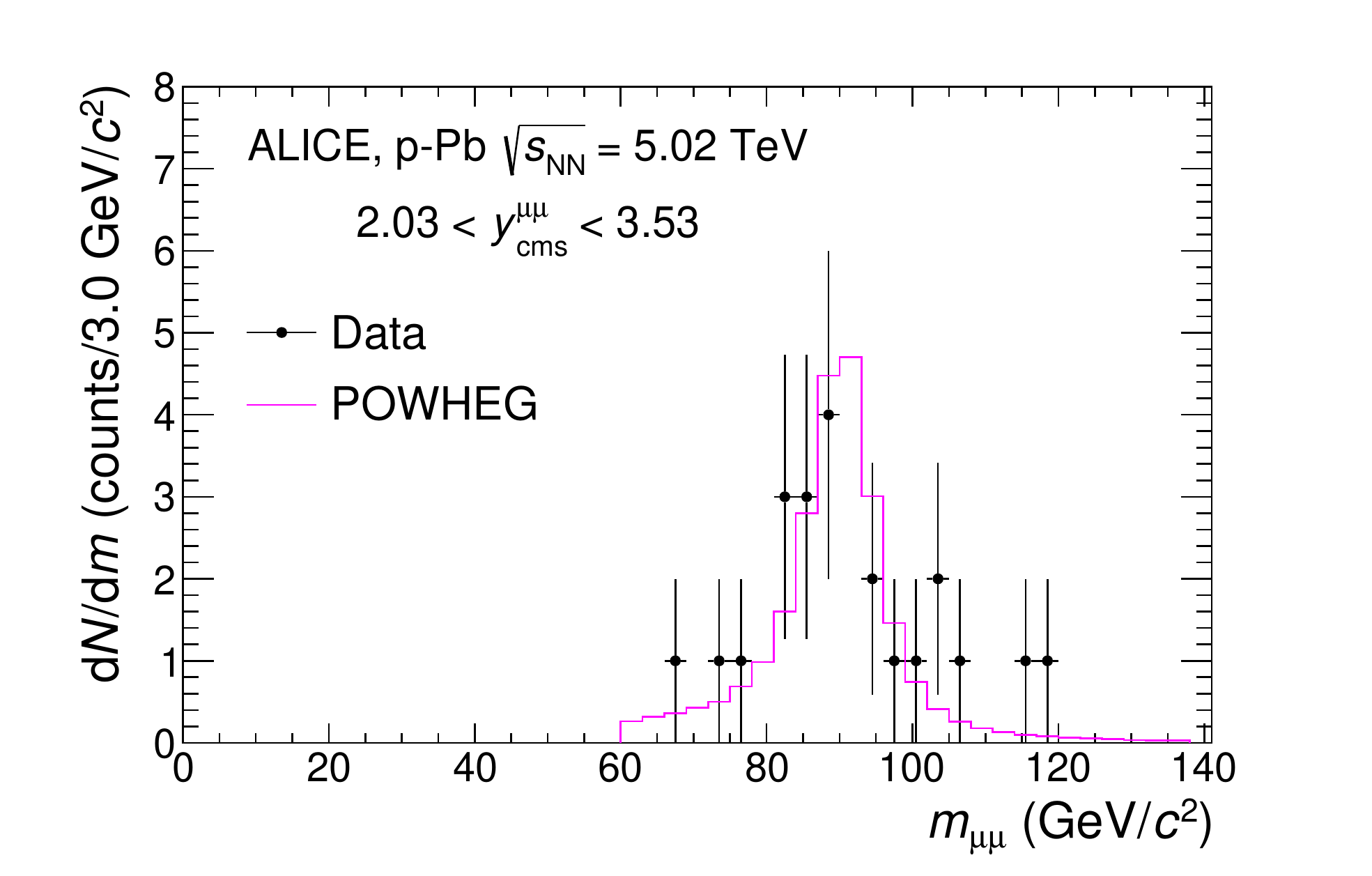}
\caption{Invariant-mass distribution of unlike-sign muon pairs with $\pt>20$\GeVc in the Pb-going (left panel) and p-going (right panel) data samples. In the p-going one, the solid line represents the distribution obtained using POWHEG simulations and normalised to the number of Z candidates in the data.}
\label{fig:invMass}
\end{figure}

Z-boson candidates are obtained by combining opposite-charge pairs of muons, selected according to the criteria described in \sect{sec:muonSelectionAndMC} and with a transverse momentum larger than 20\GeVc.
This condition reduces the contribution of lower mass resonances and of the semi-leptonic decay of charm and beauty hadrons.
It was verified that relaxing the requirement on the minimum \pt of the muon to 10\GeVc does not introduce any additional unlike-sign dimuon pair with $\invmass > 40$\GeVcc.
The resulting invariant-mass distribution is shown in \fig{fig:invMass}.
There are 2 (22) candidates with $\invmass > 60$\GeVcc reconstructed in the Pb-going (p-going) period.

For the p-going data sample, where the number of dimuons is larger, the distribution is compared with expectations from the POWHEG MC simulations described in~\sect{sec:muonSelectionAndMC}.
The results are shown in the right panel of~\fig{fig:invMass}.

The contribution to the invariant-mass distribution from combinatorial background can be estimated using the like-sign dimuon distribution: no candidates were found in the region $60 < \invmass < 120$\GeVcc.
A 0.1\% upper limit for this contribution is obtained by extrapolating the like-sign dimuon distribution at low mass ($\invmass<20$\GeVcc) to the region of interest.
Contributions from other physics processes, like the semileptonic decays of $\mathrm{c\overline{c}}$, $\mathrm{b\overline{b}}$ and $\mathrm{t\overline{t}}$ pairs and the muonic decay of $\tau$ pairs is estimated to be less than 0.7\% (0.4\%) for the p-going (Pb-going) data taking period.
Those estimations were done using MC simulations (PYTHIA 6.425 for the first process and POWHEG for the others).
Since no background events are expected, the number of Z candidates is  obtained by counting the entries in the invariant-mass distributions of opposite-charge muon pairs of \fig{fig:invMass}.

The measured number of candidates is corrected by the \acceff evaluated with simulations.
The \acceff is estimated as the ratio of the number of reconstructed Z bosons with the same analysis cuts used in data to the number of generated ones with $-4<\eta<-2.5$ and $\pt^{\mu}>20$\GeVc.
An invariant mass cut of $60 < \invmass<120$\GeVcc is applied to both reconstructed and generated Z bosons.
The resulting \acceff is 78\% (61\%) for the p-going (Pb-going) data taking period, with a relative systematic uncertainty of 1\% (2\%).
The lower \acceff value in the Pb-going configuration is due to a smaller detector efficiency in the corresponding data-taking period.
The uncertainty accounts for the difference from the values obtained with a simulation based on the residual misalignment and that based on the data-driven alignment.
The systematic uncertainties are summarised in \tab{tab:zSyst}.

\begin{table}[ht]
 \centering
 \begin{tabular}{|l|c|}
 \hline
 Background contamination & $<1\%$ \\
 \hline
 Tracking efficiency & 4\% (p-going) ~~ 6\% (Pb-going) \\
 Trigger efficiency & 2\% \\
 Tracker/trigger matching & 1\% \\
 Alignment & 1\% (p-going) ~~ 2\% (Pb-going)  \\
 \hline
 \fnorm & 1\% \\
 MB cross section & 3.3\% \\
 \hline
 \end{tabular}
 \caption{Summary of systematic uncertainties for Z-boson analysis.}
 \label{tab:zSyst}
\end{table}

\subsection{W-boson analysis}\label{sec:wAnalysis}
At transverse momenta higher than 10\GeVc, the main contributions to the inclusive \pt distribution of muons are the decays of W bosons, the dimuon decays of Z bosons and the muon decays of heavy-flavoured hadrons.
The number of muons from W decays can be extracted from the inclusive \pt spectrum before \acceff corrections through a fit procedure based on MC template descriptions of these three main components:
\begin{equation} \label{eq:fitFull}
f(\pt) = \fitNbkg \fsub{bkg}(\pt) + \fitNw (\fsub{\wmu}(\pt) + R \fsub{\zmu}(\pt))
\end{equation}
where {\fsub{bkg}}, {\fsub{\wmu}} and {\fsub{\zmu}} are the MC templates for muons from heavy-flavoured hadrons, W-boson and Z-boson decays, respectively.
The number of muons from heavy-flavour decays (\fitNbkg) and the number of muons from W decays (\fitNw) are free parameters, while the ratio ($R$) of the number of muons from Z decays and that from W decays is fixed from MC simulations using POWHEG.
It was verified that these calculations well describe the measured Z boson production in the dimuonic decay channel, described in the previous section.
The contribution of muons from heavy-flavour decays was simulated using as input the QCD calculations in the Fixed-Order Next-to-Leading-Log (FONLL) approach~\cite{Cacciari:2012ny}, which are found to provide a good description of data in pp collisions.
The calculations were obtained using the CTEQ6.6 parton distribution functions~\cite{Nadolsky:2008zw}, without accounting for any nuclear modification.
Such modifications, however, mainly affect the production at low transverse momenta, with a negligible effect in the shape of the \pt distribution in the region of interest for this study~\cite{Abelev:2014hha}.
The templates for muons from the decay of W and Z bosons were obtained with MC simulations based on POWHEG.
The detector response is included in all simulations.

\begin{figure}[t]
\centering
\includegraphics[width=0.48\textwidth]{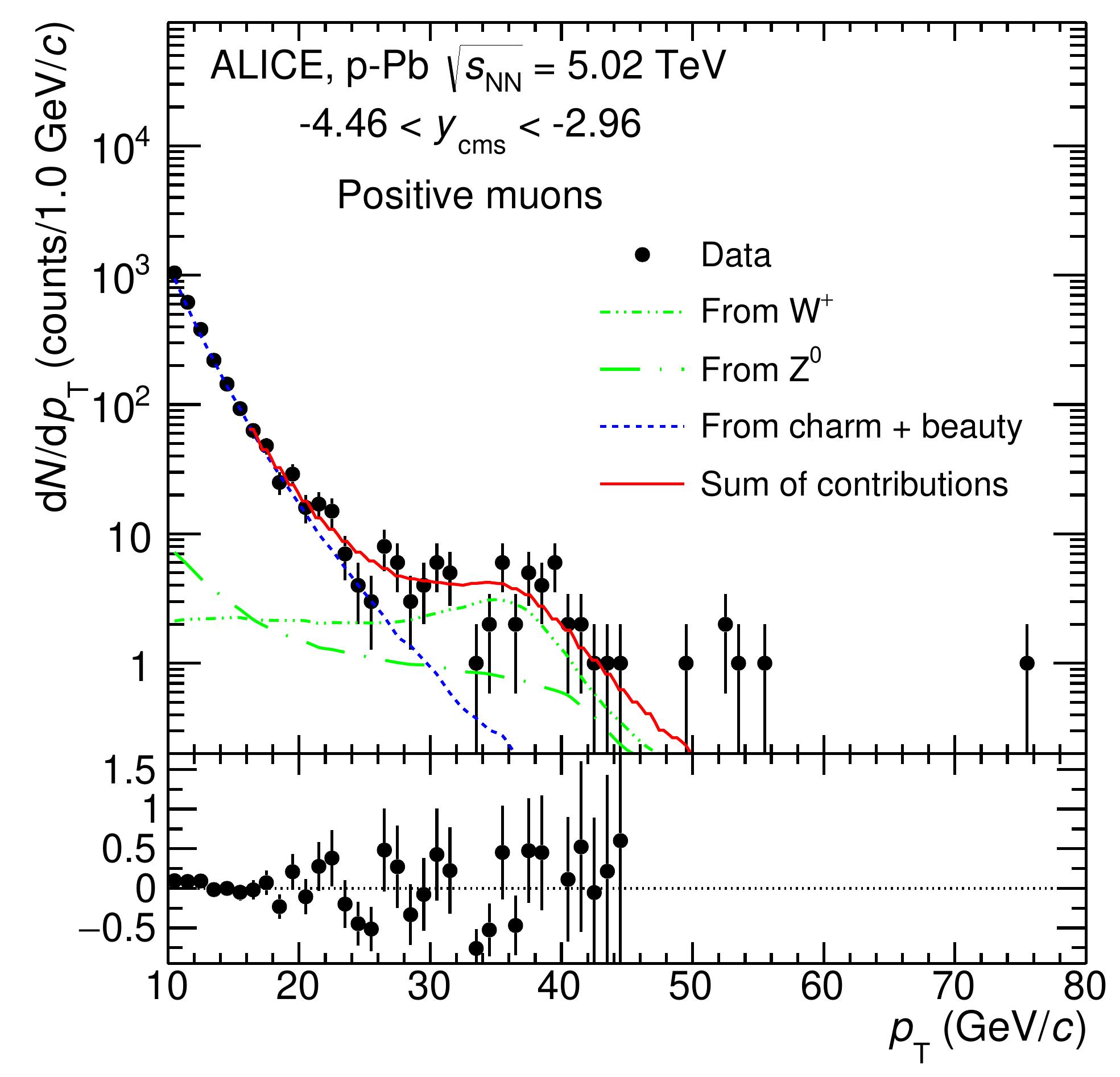}
\includegraphics[width=0.48\textwidth]{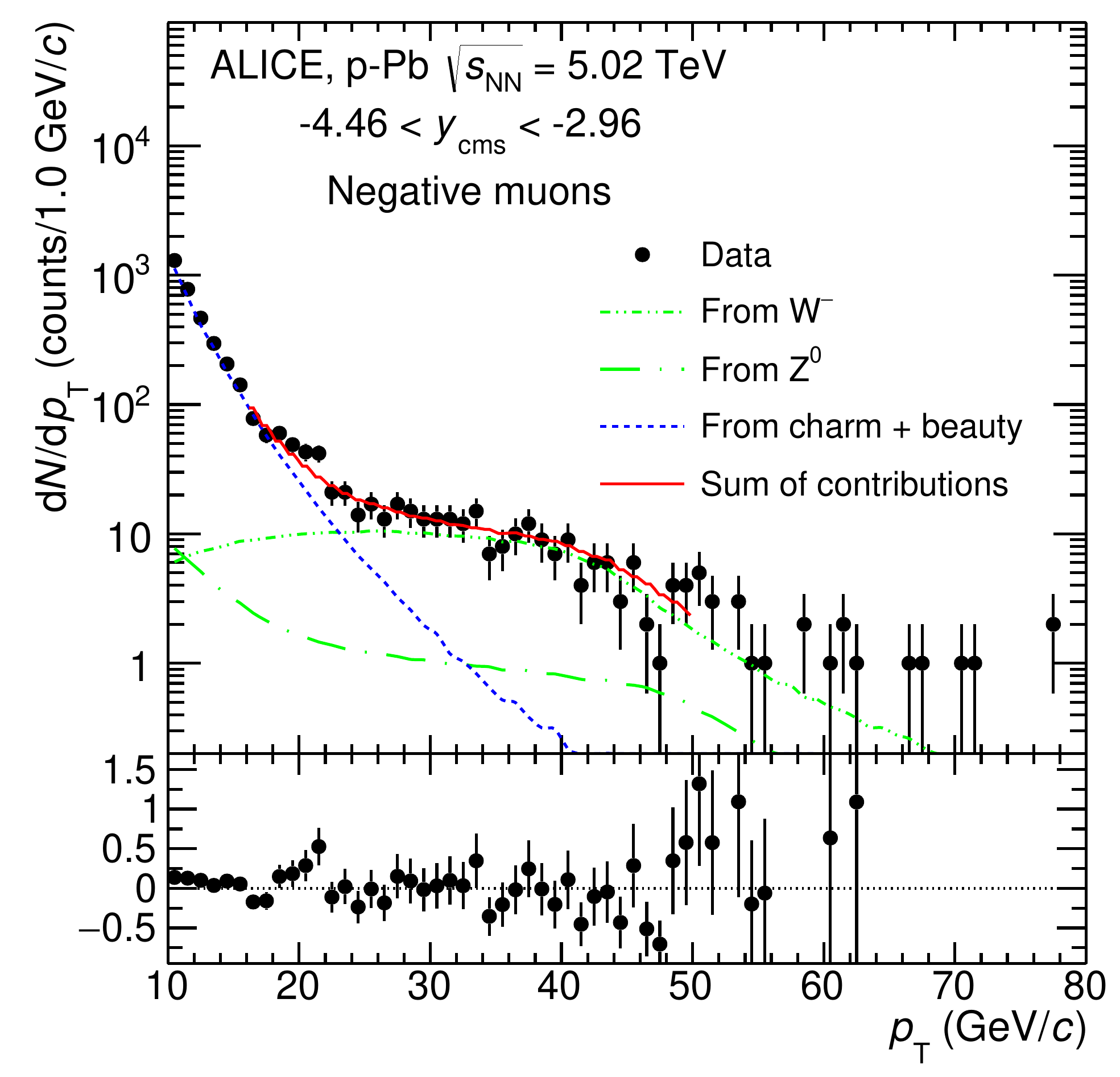}

\includegraphics[width=0.48\textwidth]{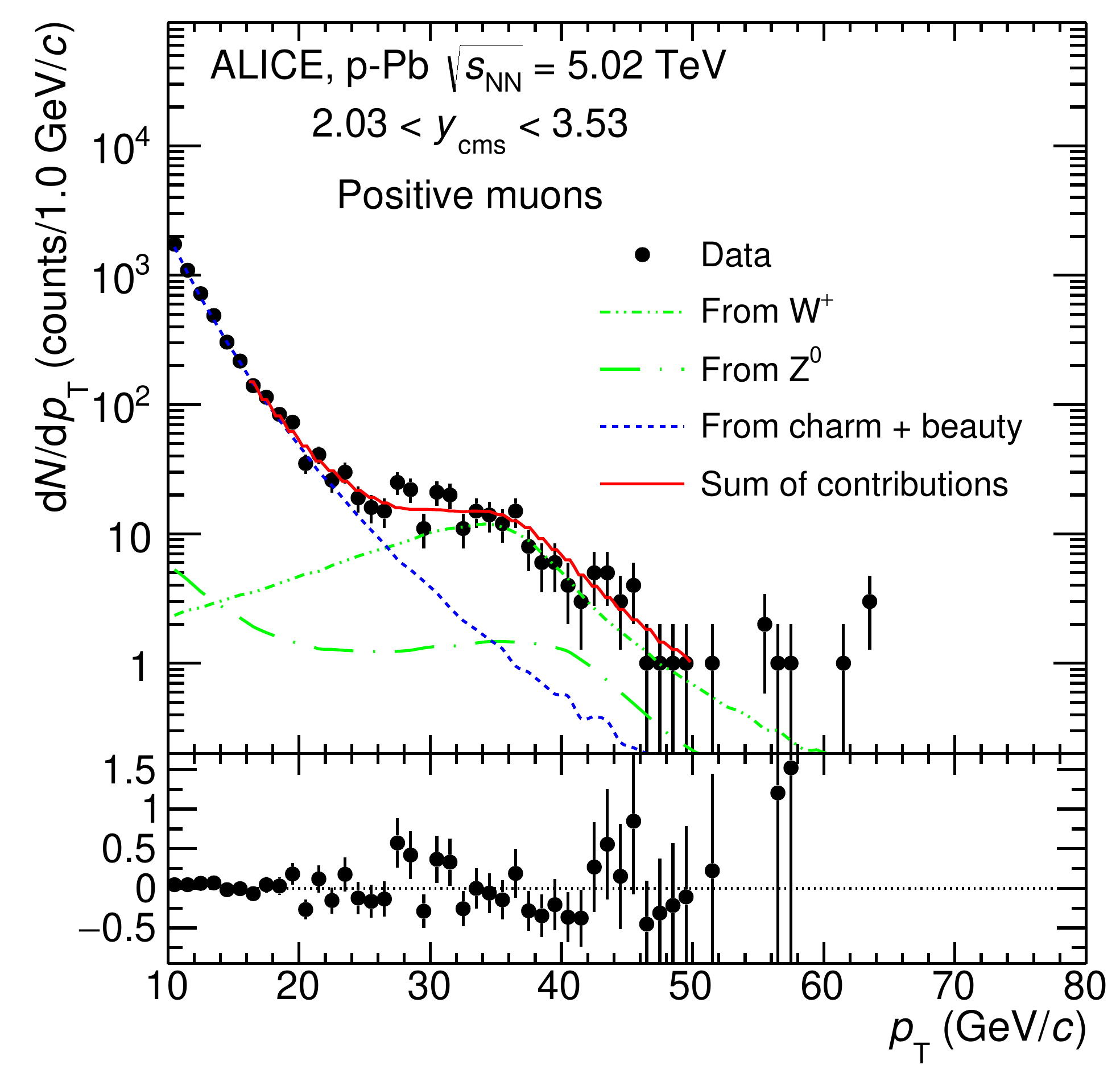}
\includegraphics[width=0.48\textwidth]{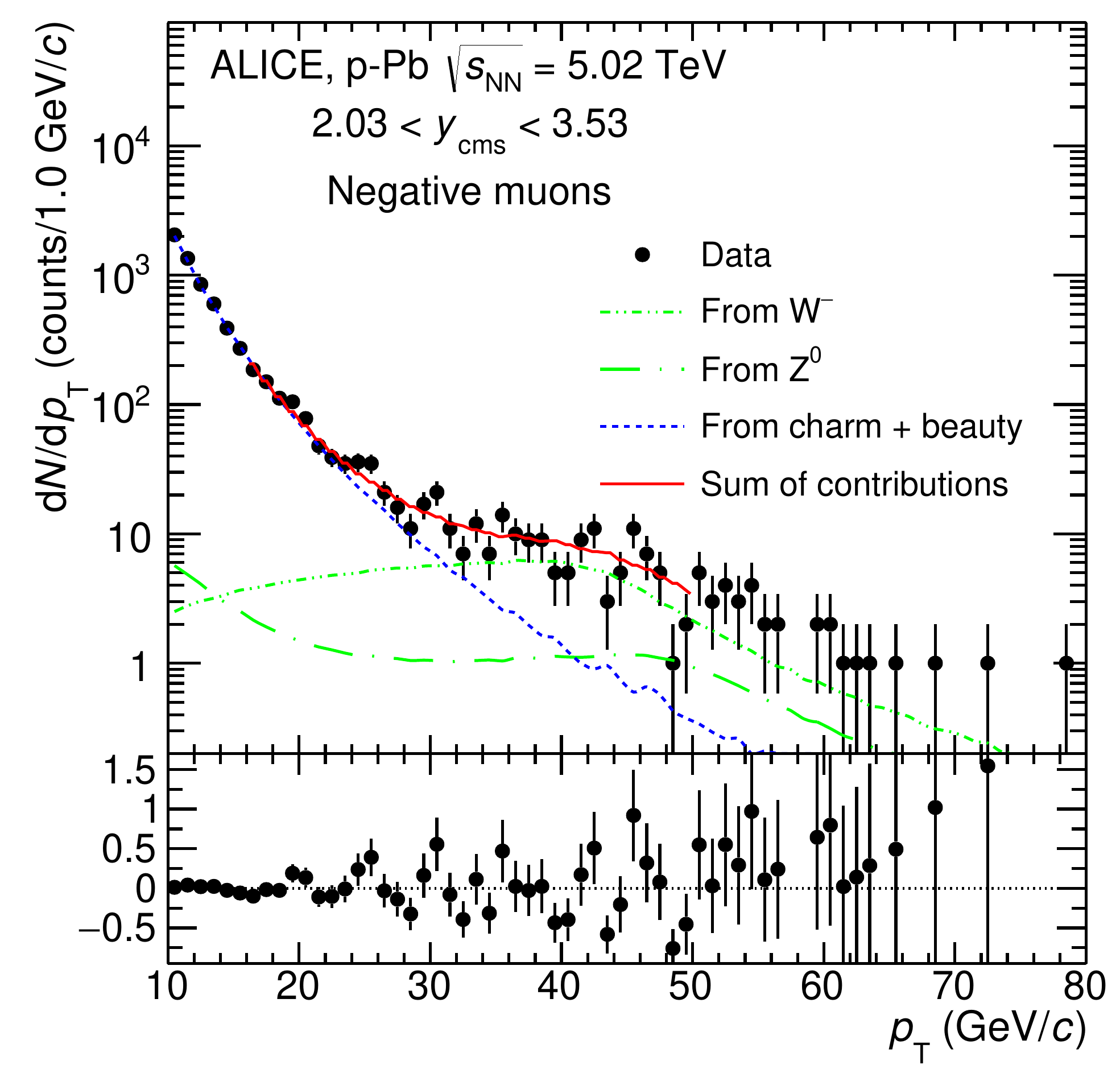}
\caption{Top panels: inclusive distribution of positive (left) and negative (right) charge muon candidates measured in the Pb-going (top) and p-going (bottom) data taking periods. The results of the MC template fit for the extraction of the \wmus[+] and \wmus[-] signal is shown.
In this case, the central value of the FONLL calculations is used for the background description while POWHEG with the CT10 PDF set paired with EPS09NLO is used for W and Z boson production.
Bottom panels: relative difference of data and the extrapolated fit results in the range $10<\pt<80$\GeVc.}
\label{fig:fitExample}
\end{figure}

The inclusive transverse momentum distributions of positive and negative muon candidates passing the selections described in~\sect{sec:muonSelectionAndMC} are fitted according to~\eq{eq:fitFull}, and the parameter \fitNw is extracted from the fit.
The MC templates are then modified as explained later on to account for the uncertainties affecting their shape and the fit is performed again, thus yielding different values of \fitNw.
The procedure is reiterated for each set of MC templates considered.
The number of muons from W decays is finally estimated as the arithmetic average of the
\fitNw extracted in each fit, while their dispersion, estimated as the Root Mean Square (RMS) of the \fitNw distribution, is used as systematic uncertainty.
An example of signal extraction for a specific set of MC templates is shown in~\fig{fig:fitExample}.

Several sources of uncertainty affecting the shape of the MC templates were taken into account.
For the background, different MC templates were obtained by varying the FONLL calculations within uncertainties.
In particular, six additional templates were produced, corresponding to the upper and lower limits of the calculations obtained by i) varying the factorisation and renormalisation scales, and considering the uncertainties on ii) the quark masses and iii) the PDFs.
For the W and Z boson production, different PDF sets were used, both at LO and NLO, in particular the CT10~\cite{Lai:2010vv} and CTEQ6~\cite{Pumplin:2002vw} paired with EPS09.
The use of different sets affects both the shapes of the templates and the cross-sections, thus resulting in a variation of the parameter $R$ in \eq{eq:fitFull}.
The stability of the fit was tested by varying the lower limit of the transverse momentum range (the upper one being mainly limited by statistics) from 15 to 17\GeVc.
Finally, the effect of the momentum resolution was accounted for by using two different sets of templates for each MC input, obtained by including in the simulations either the tracking chamber residual misalignment or the data-driven method discussed in~\sect{sec:muonSelectionAndMC}.
The contamination to the positive muon spectrum of negative muons with mis-identified charge sign (and viceversa) is estimated to be smaller than 1\% for $\pt>10$\GeVc.
The contamination depends on the \pt, but the resulting variation of the yields in the \pt range of the fit is found to be smaller than the variation of the shape of the templates obtained with different descriptions of the alignment.

The number of muons from W-boson decays is then corrected for the detector acceptance and efficiency.
The values of \acceff integrated over $\pt^{\mu}>10$~\GeVc are 89\% for $\mu^+$ and 88\% for $\mu^-$ in the p-going period and of 77\% for $\mu^+$ and 75\% for $\mu^-$ in the Pb-going period.
The lower \acceff value in the Pb-going configuration is due to a smaller detector efficiency in the corresponding data-taking period.
A difference of 1\% in the values is observed when using the data-driven method for the description of the alignment in the simulations instead of the residual misalignment.
This value is taken as the systematic uncertainty.

All systematic uncertainties are summarised in \tab{tab:wSyst}.

\begin{table}[t]
 \centering
 \begin{tabular}{|l|c|}
  \hline
 Signal extraction & 2 -- 6\% \\
 ~~- vs centrality & 5 -- 15\% \\
 \hline
 Tracking efficiency (c) & 2\% (p-going) ~~ 3\% (Pb-going)\\
 Trigger efficiency (c) & 1\% \\
 Tracker/trigger matching (c) & 0.5\% \\
 Alignment (c) & 1\% \\
 \hline
 \fnorm (c) & 1\% \\
 MB cross section (c) & 3.3\%  \\
 \hline
 Pile-up & 1 - 3 \% \\
 $\langle \ncmult \rangle$ & 2 -- 8\% \\
 \hline
\end{tabular}
\caption{Summary of systematic uncertainties for W-boson analysis. The uncertainties that are correlated between measurements in different centrality bins are indicated with (c).}
\label{tab:wSyst}
\end{table}

\section{Results}\label{sec:results}
\begin{figure}[t]
\centering
\includegraphics[width=0.48\textwidth]{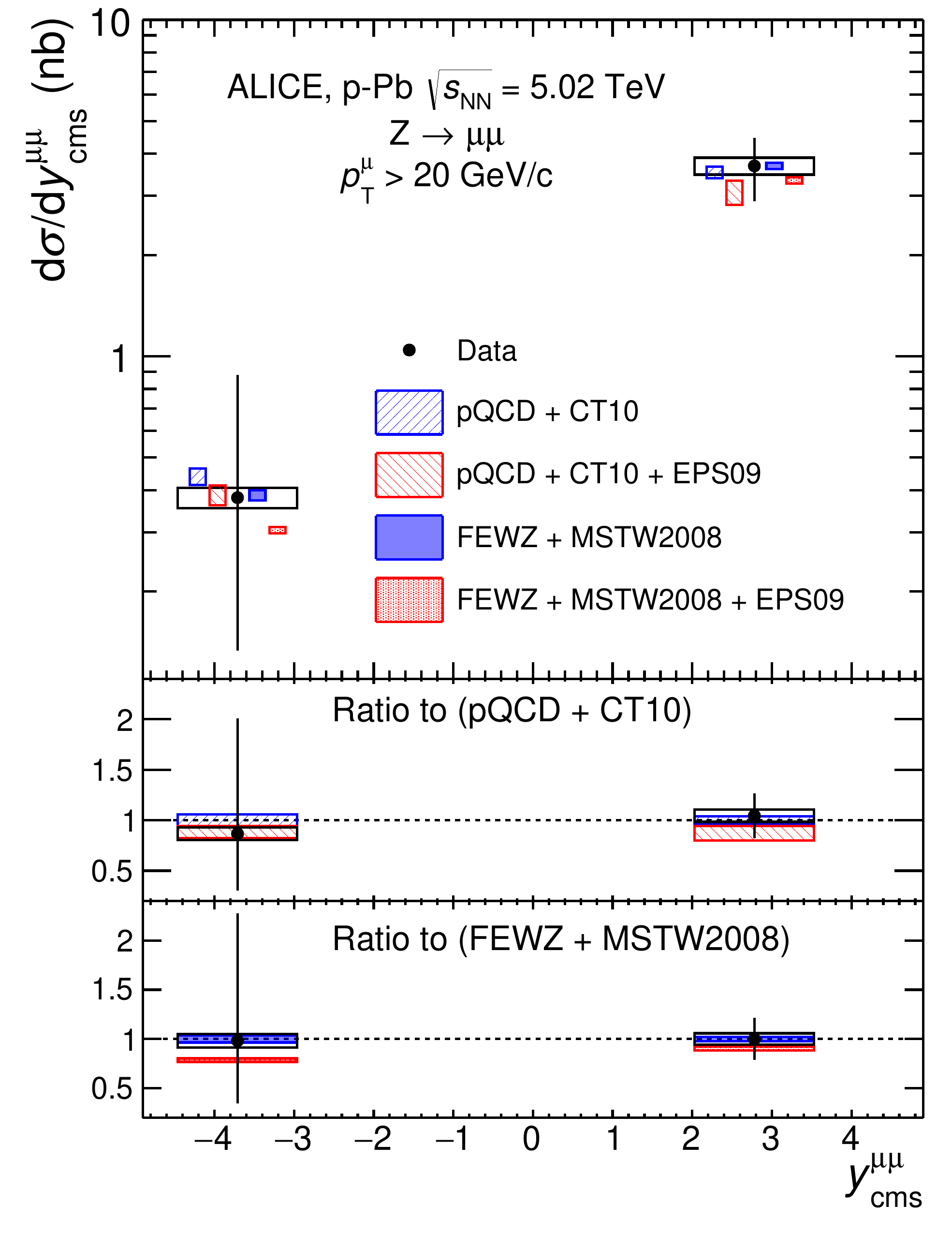}
\caption{Z-boson production cross section in the dimuon decay channel at backward and forward rapidities measured in \pPb collisions at \snn = 5.02~TeV. The vertical error bars (open boxes) represent the statistical (systematic) uncertainties. The horizontal width of the boxes corresponds to the measured rapidity range. The results are compared with theoretical calculations~\cite{Paukkunen:2010qg,Gavin:2010az} performed both with and without including the nuclear modification of the parton distribution functions. In the top panel, the calculations are shifted along the rapidity axis to improve the visibility. The middle (bottom) panel shows the data and pQCD (FEWZ) calculations divided by the pQCD (FEWZ) calculations without nuclear modification of the PDFs.}
\label{fig:zCrossSection}
\end{figure}

The Z-boson production cross section in the dimuon decay channel with $\pt^\mu>20$\GeVc and $60<\invmass<120$\GeVcc is shown in \fig{fig:zCrossSection}.
The vertical bars represent the statistical uncertainties while the open boxes are the systematic ones.
The cross section at backward rapidity is estimated from two reconstructed Z boson candidates (see left panel of~\fig{fig:invMass}).
In this case, the statistical uncertainty is defined as the 68\% confidence interval assuming a Poisson distribution for the number of Z bosons.
Moreover, an upper limit was also calculated, whose value is of 1.75~nb at a 95\% confidence level.
The results are compared with NLO and NNLO theoretical calculations both with and without including the nuclear modification of the parton distribution functions.
The NLO pQCD calculations~\cite{Paukkunen:2010qg} (blue hatched boxes) are obtained using the CT10~\cite{Lai:2010vv} PDF, while the NNLO calculations with FEWZ~\cite{Gavin:2010az} (blue filled boxes) use the MSTW2008 NNLO~\cite{Martin:2009iq} PDF set.
Both calculations describe the data within uncertainties.
The corresponding calculations with the EPS09NLO parameterisation of the nuclear modification of the parton distribution functions are shown as hatched and filled red boxes, respectively.
The nuclear effect results in a small reduction of the cross section, in particular at forward rapidities where lower Bjorken-$x$ values of the Pb nucleons are probed.
The effect, however, is small and the measurement is compatible with both calculations within uncertainties.

\begin{figure}[!t]
\centering
\includegraphics[width=0.6\textwidth]{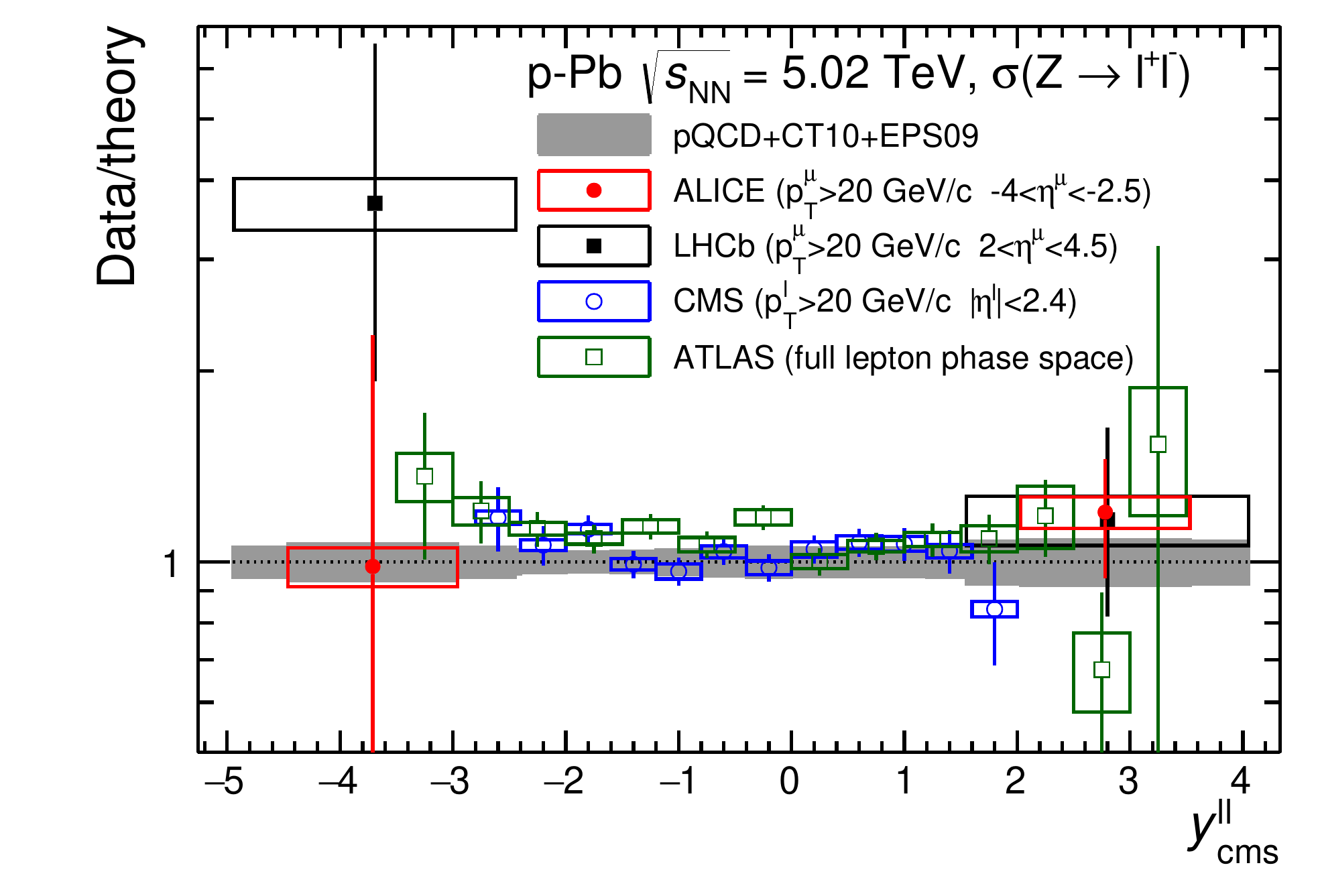}
\caption{
Ratio of data over theoretical calculations for the Z-boson production cross section measured by the ALICE, LHCb~\cite{Aaij:2014pvu}, ATLAS~\cite{Aad:2015gta} and CMS~\cite{Khachatryan:2015pzs} experiments. The LHCb points have been shifted by +0.02 units of rapidity for better visibility.
The ATLAS cross sections are measured in a slightly smaller invariant mass range ($66 < m_{ll} < 116$\GeVcc) compared to the other experiments ($60 < m_{ll} < 120$\GeVcc).
The luminosity uncertainties of 2.7\% for ATLAS and 3.5\% for CMS are not shown.
The pQCD calculations are obtained with the CT10 PDF set and with the EPS09NLO parameterisation of the nuclear modifications.}
\label{fig:LHCreviewZ}
\end{figure}

The Z-boson production cross section was measured in \pPb collisions at \snn = 5.02~TeV by the ATLAS and CMS experiments at mid-rapidity~\cite{Aad:2015gta,Khachatryan:2015pzs} and by the LHCb experiment at forward and backward rapidities~\cite{Aaij:2014pvu}.
The LHCb measurement is performed in a wider pseudorapidity interval ($2<\eta<4.5$) compared to ALICE, but on a data sample with a smaller integrated luminosity.
Figure~\ref{fig:LHCreviewZ} shows the cross section measurements of the four LHC experiments, each divided by the corresponding NLO pQCD expectation including the nuclear modification of the PDFs~\cite{Paukkunen:2010qg}: the calculations are found to describe all data.
It is worth noting, however, that none of the experiments can exclude the calculations without nPDFs.

\begin{figure}[t]
\centering
\includegraphics[width=0.48\textwidth]{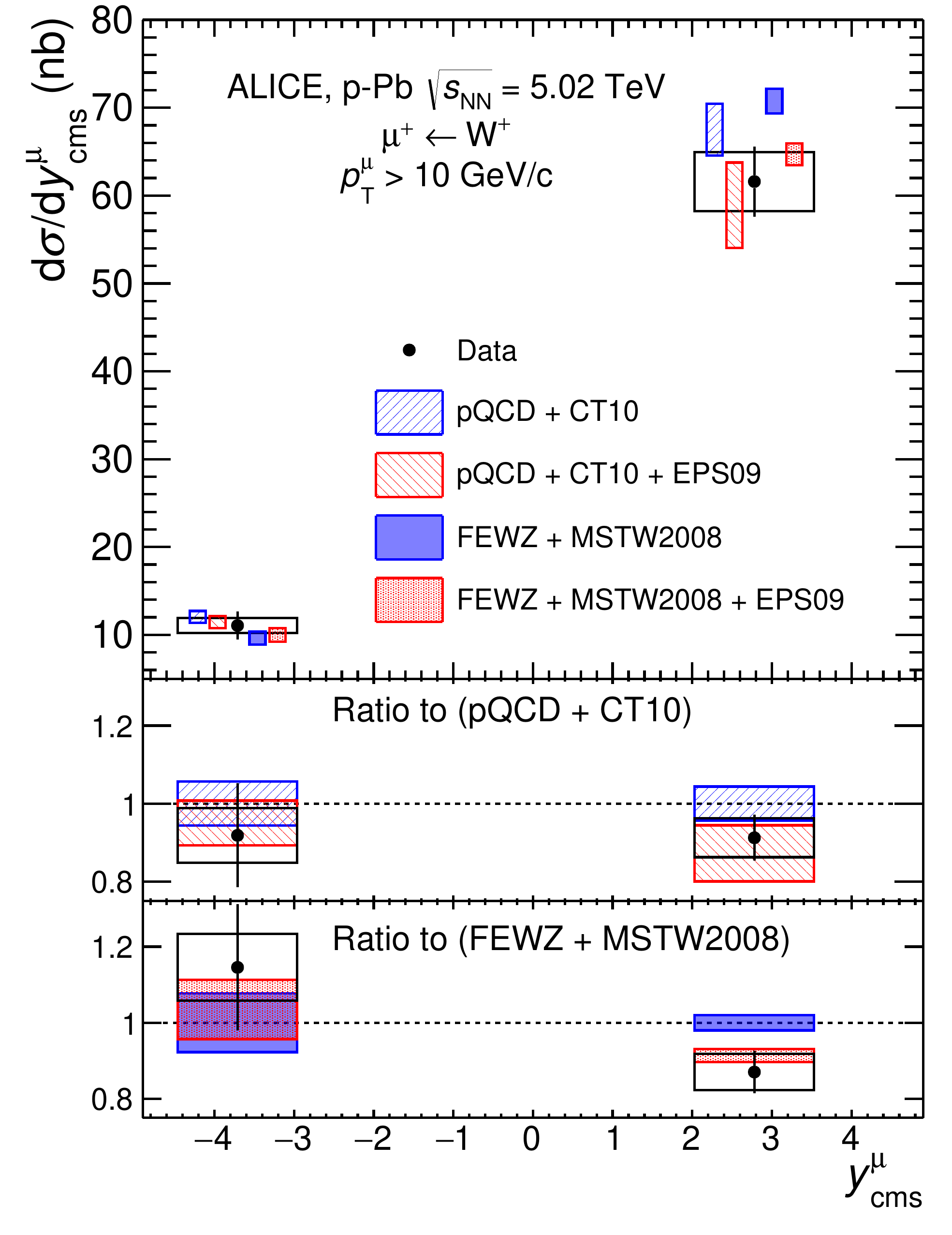}
\includegraphics[width=0.48\textwidth]{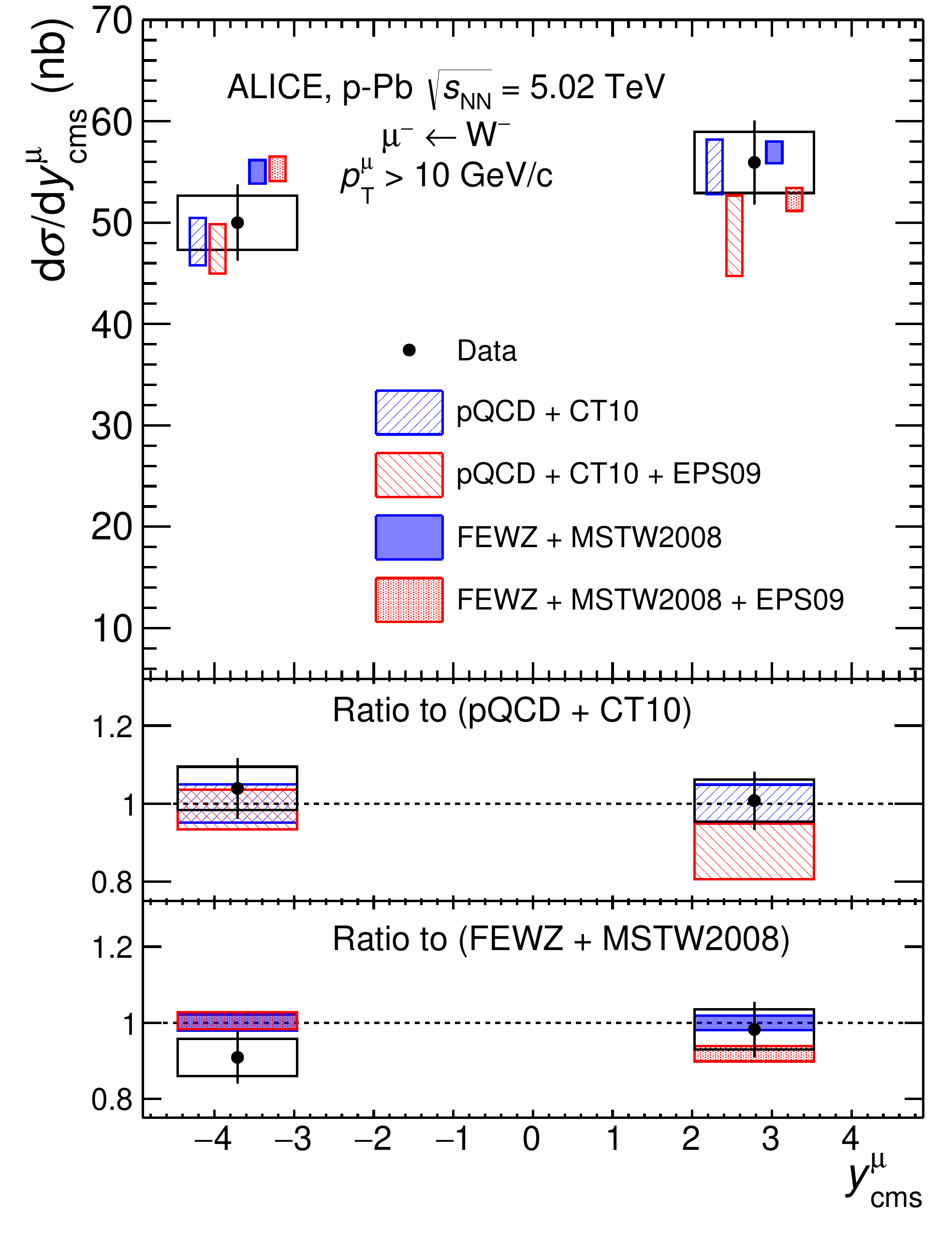}
\caption{Left (right) panel: cross section of $\mu^+$ ($\mu^-$) from \wBoson[+] (\wBoson[-]) boson decays at backward and forward rapidities measured in \pPb collisions at \snn = 5.02~TeV. The vertical error bars (open boxes) represent the statistical (systematic) uncertainties. The horizontal width of the boxes corresponds to the measured rapidity range. The results are compared with theoretical calculations~\cite{Paukkunen:2010qg,Gavin:2010az} performed both with and without including the nuclear modification of the parton distribution functions. In the top panels, the calculations are shifted along the rapidity axis to improve the visibility. The middle (bottom) panel shows the data and pQCD (FEWZ) calculations divided by the pQCD (FEWZ) calculations without nuclear modification of the PDFs.}
\label{fig:wCrossSection}
\end{figure}

The cross sections of muons from \wBoson[+] and \wBoson[-] boson decays with $\pt^\mu>10$\GeVc measured at forward and backward rapidities in \pPb collisions at \snn = 5.02~TeV are shown in the left and right panels of \fig{fig:wCrossSection}, respectively.
The vertical bars represent the statistical uncertainties while the open boxes are the systematic ones.
The smaller cross-section of positive W bosons at backward rapidity is the combined effect of the parity violation of the weak interaction, which only couples left-handed fermions with right-handed anti-fermions, and of the helicity conservation in the leptonic decay.
This results in an anisotropic emission of the muons.
In particular, the $\mu^-$ is preferably emitted in the same direction of the \wBoson[-], while the $\mu^+$ is emitted in the opposite direction with respect to the \wBoson[+]~\cite{ConesaDelValle:2007dza}.
This implies that the $\mu^+$ measured in $-4.46 < \ycms < -2.96$ mainly comes from the decay of \wBoson[+] at even more backward rapidities, where the production cross-section rapidly decreases.

The results are compared with the analogous model calculations used to describe the Z-boson production.
The NLO pQCD calculations with CT10 parton distribution functions (blue hatched boxes) and the NNLO calculations with FEWZ with the MSTW2008 PDF set (blue filled boxes) both describe the data within uncertainties.
The inclusion of a parameterisation of the nuclear modification of the parton distribution function in the calculations (red hatched boxes for pQCD and red filled boxes for FEWZ) results in a slightly lower value of the cross section, especially at forward rapidity.
This variation, however, is of the same order as the uncertainties in the theoretical calculations, thus limiting the discriminating power of the cross section alone.

\begin{figure}[t]
\centering
\includegraphics[width=0.48\textwidth]{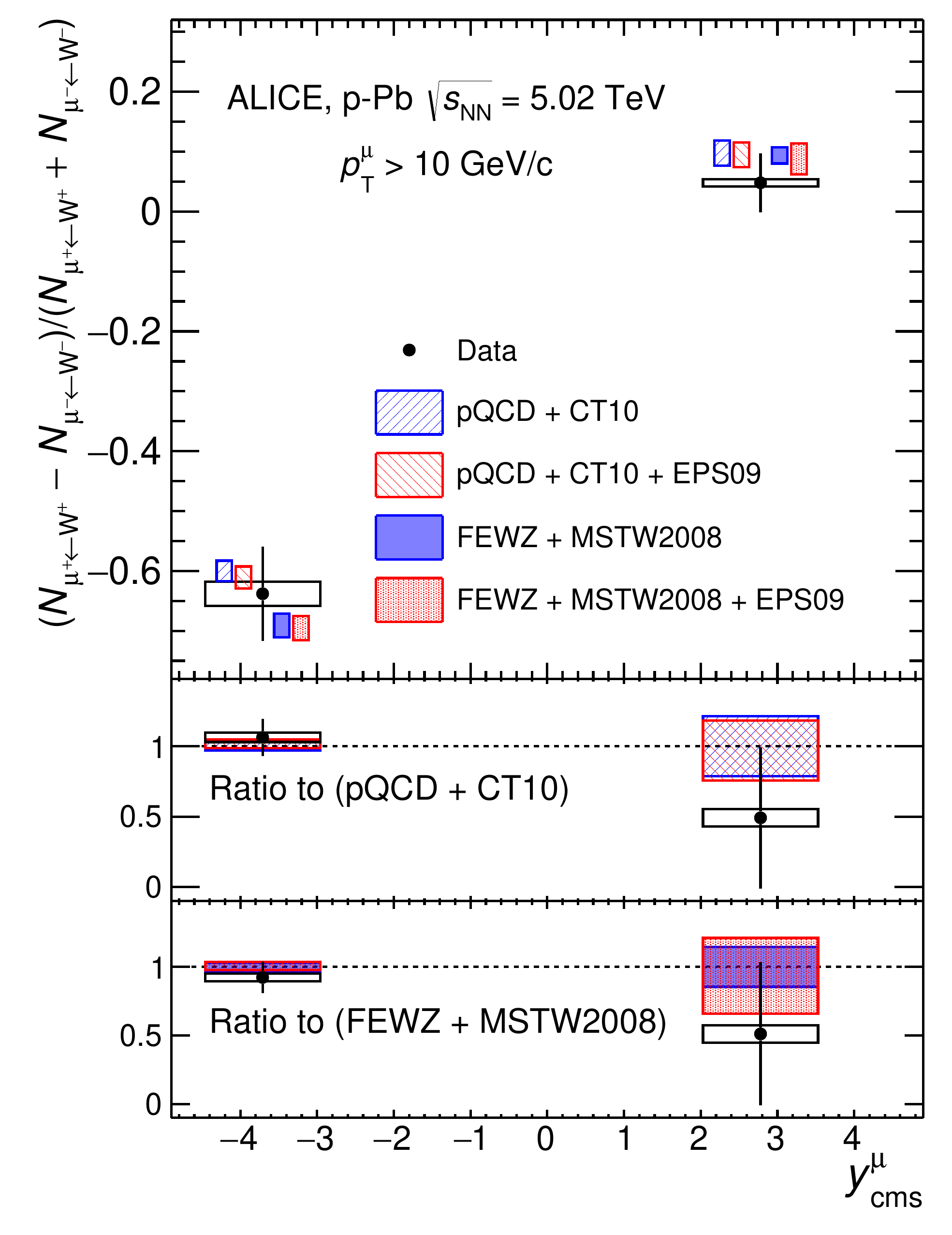}
\caption{Lepton charge asymmetry of muons from W-boson decays at backward and forward rapidities measured in \pPb collisions at \snn = 5.02~TeV. The vertical error bars (open boxes) represent the statistical (systematic) uncertainties. The horizontal width of the boxes corresponds to the measured rapidity range. The results are compared with theoretical calculations~\cite{Paukkunen:2010qg,Gavin:2010az} performed both with and without including the nuclear modification of the parton distribution functions. In the top panel, the calculations are shifted along the rapidity axis to improve the visibility. The middle (bottom) panel shows the data and pQCD (FEWZ) calculations divided by the pQCD (FEWZ) calculations without nuclear modification of the PDFs.}
\label{fig:wChargeAsymmetry}
\end{figure}

The asymmetry in the production of the \wBoson[+] and \wBoson[-] bosons can be used to gain sensitivity in the study of the nuclear modification of the PDFs~\cite{Khachatryan:2015hha}.
Part of the theoretical uncertainties, such as those on the factorization and renormalization scale that are of the order of 5\%, and the experimental uncertainties on the tracking and trigger efficiency, normalisation factors and MB cross section, whose quadratic sum amounts to 4.3\% (4.8\%) in the p-going (Pb-going) period, cancel when measuring the relative yield of muons from \wBoson[+] and \wBoson[-] decays.
Figure~\ref{fig:wChargeAsymmetry} shows the lepton charge asymmetry, which is defined as:
\begin{equation}\label{eq:chargeAsymmetry}
\frac{\nsub{\wmus[+]}-\nsub{\wmus[-]}}{\nsub{\wmus[+]}+\nsub{\wmus[-]}}
\end{equation}
where $\nsub{\wmus[+]}$ and $\nsub{\wmus[-]}$ are the yields of muons from, respectively, the \wBoson[+] and \wBoson[-] decays, corrected by the detector acceptance and efficiency.
The relative systematic uncertainties in the pQCD and FEWZ calculations are strongly reduced in the ratio.
However, the model results with and without nuclear modification are very similar in this kinematic range, and the measurement cannot discriminate between them.

\begin{figure}[!t]
\centering
\includegraphics[width=0.48\textwidth]{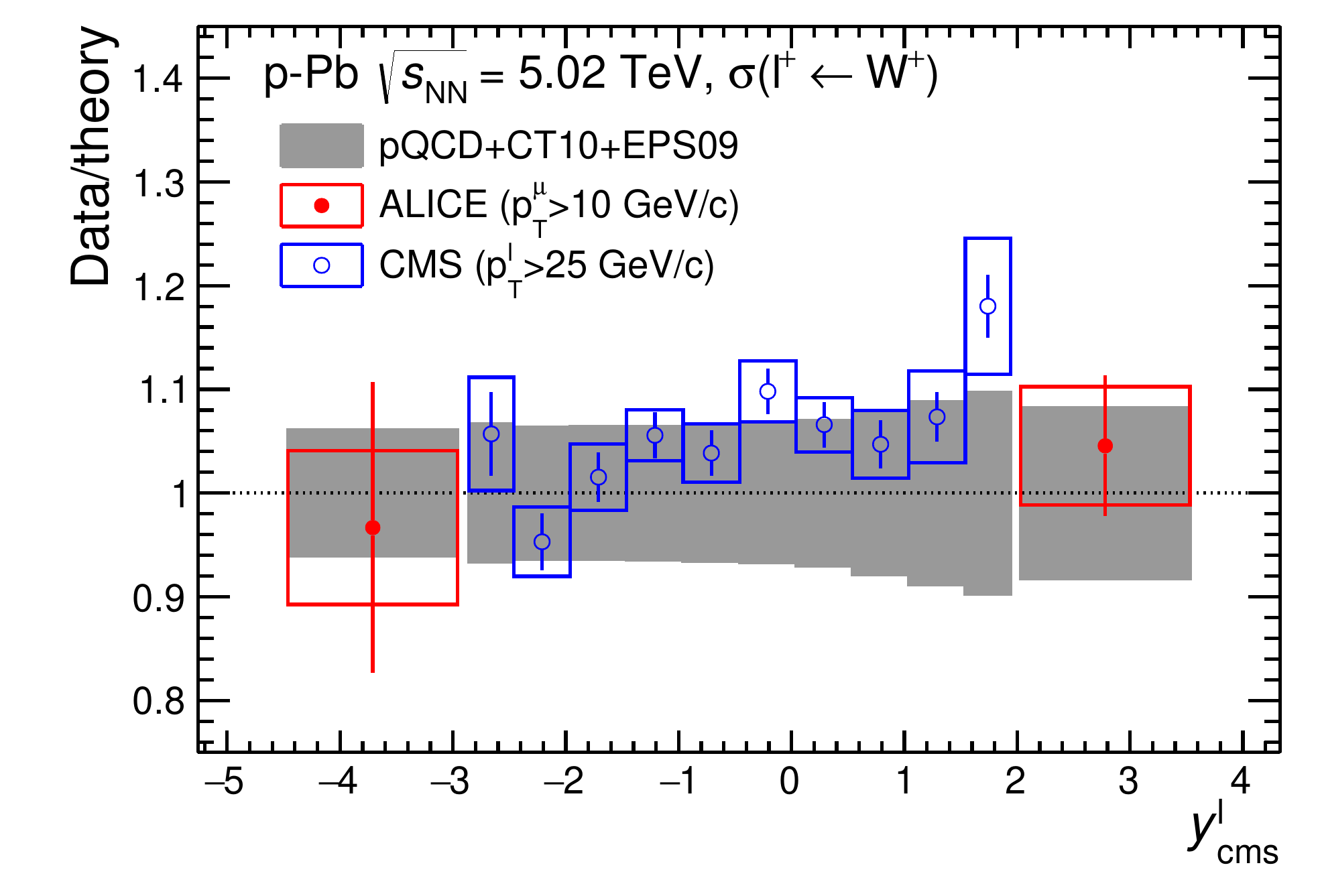}
\includegraphics[width=0.48\textwidth]{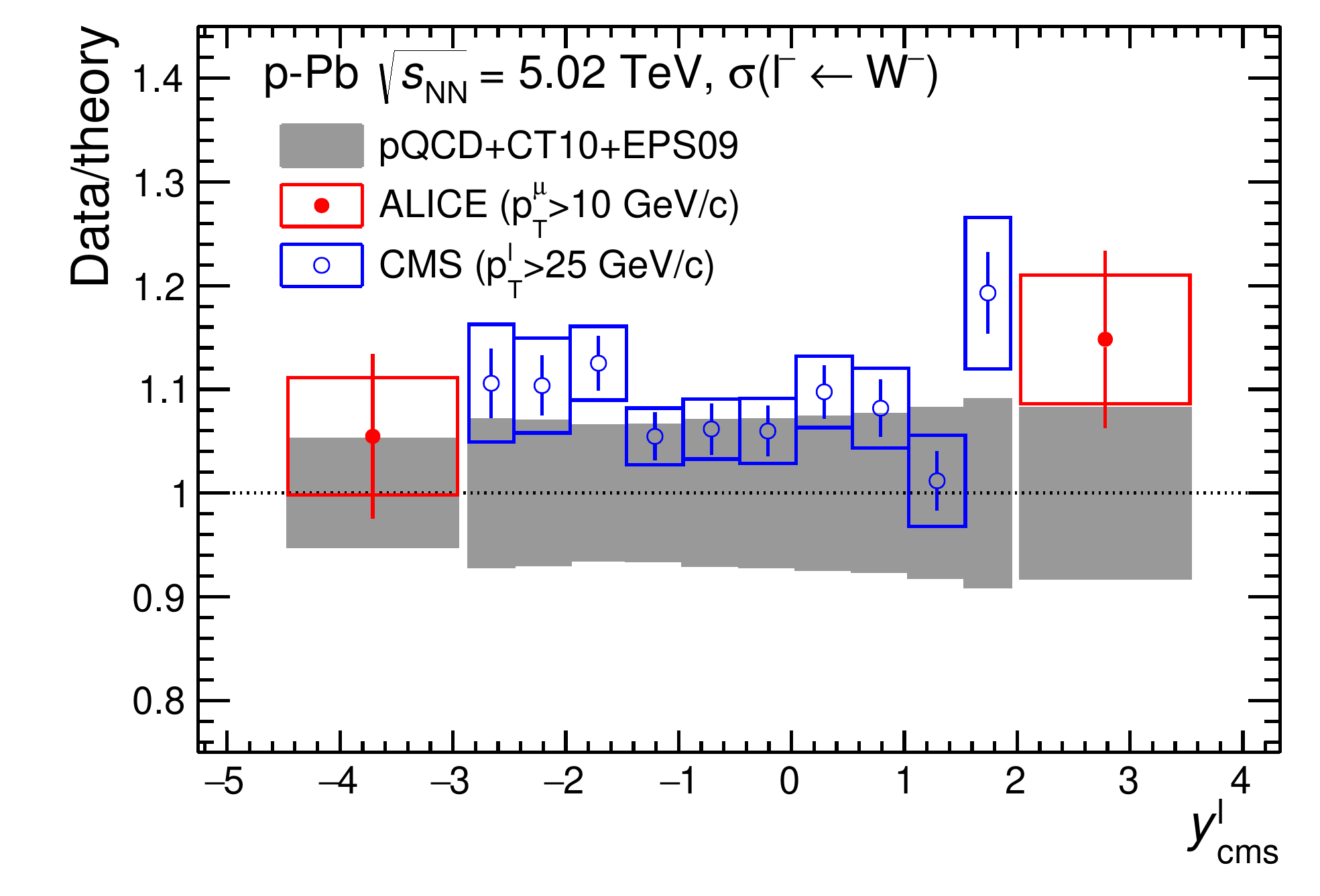}
\caption{Ratio of data over theoretical calculations for the production cross section of positive (left panel) and negative (right panel) muons and leptons from W-boson production measured by the ALICE and CMS experiments~\cite{Khachatryan:2015hha}, respectively. The luminosity uncertainty of 3.5\% for CMS is not shown. The pQCD calculations are obtained with CT10 NLO PDF set and with the EPS09NLO parameterisation of the nuclear modifications.}
\label{fig:LHCreviewW}
\end{figure}

The production of electrons and muons from W-boson decays was measured at mid-rapidity in \pPb collisions at \snn = 5.02~TeV by the CMS experiment~\cite{Khachatryan:2015hha}.
The cross section results, each divided by the corresponding NLO pQCD expectation including nuclear modification of the PDFs, are shown together with the analogous ALICE results in \fig{fig:LHCreviewW}: the calculations are found to describe data over the full explored rapidity interval.

\begin{figure}[t]
\centering
\includegraphics[width=0.48\textwidth]{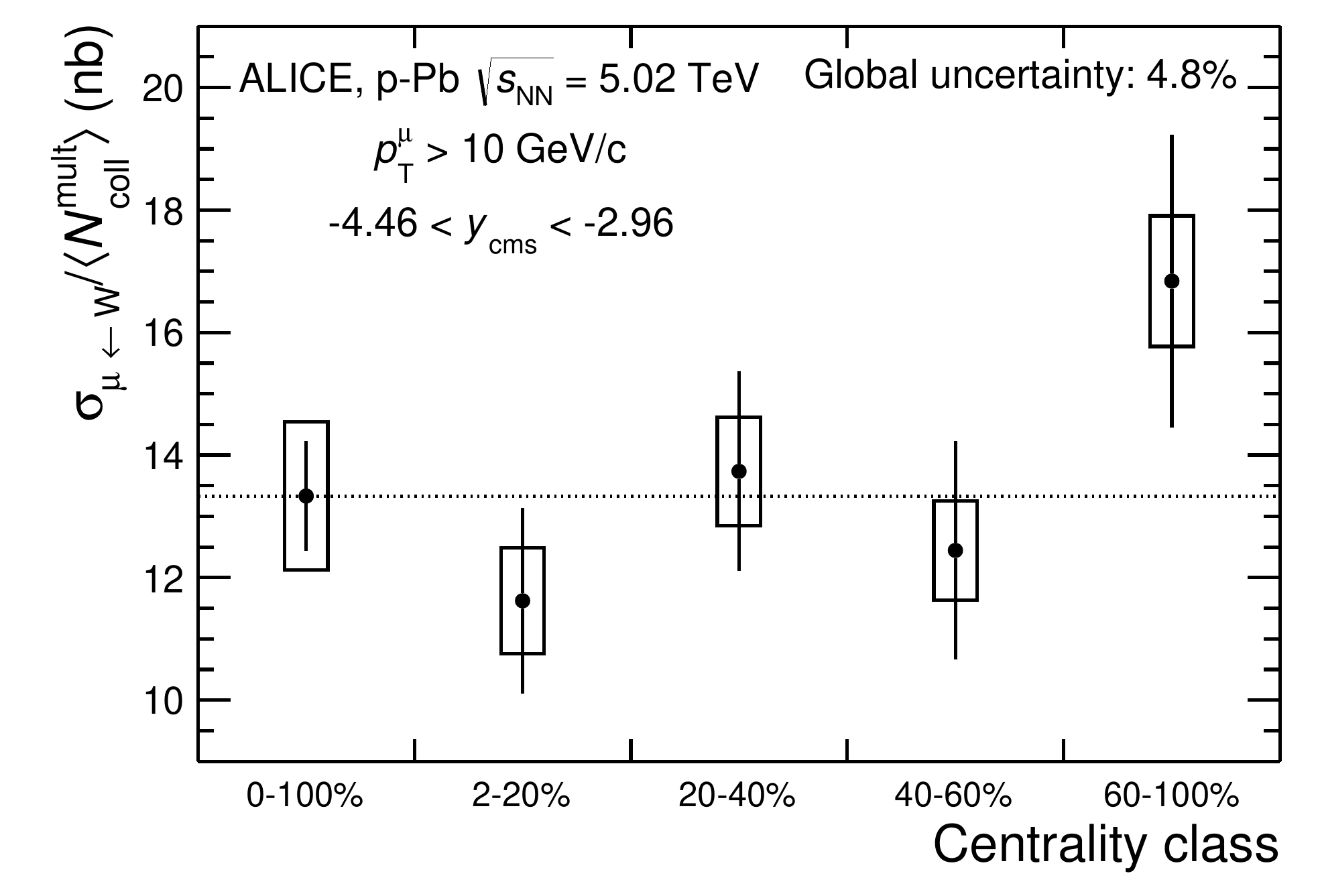}
\includegraphics[width=0.48\textwidth]{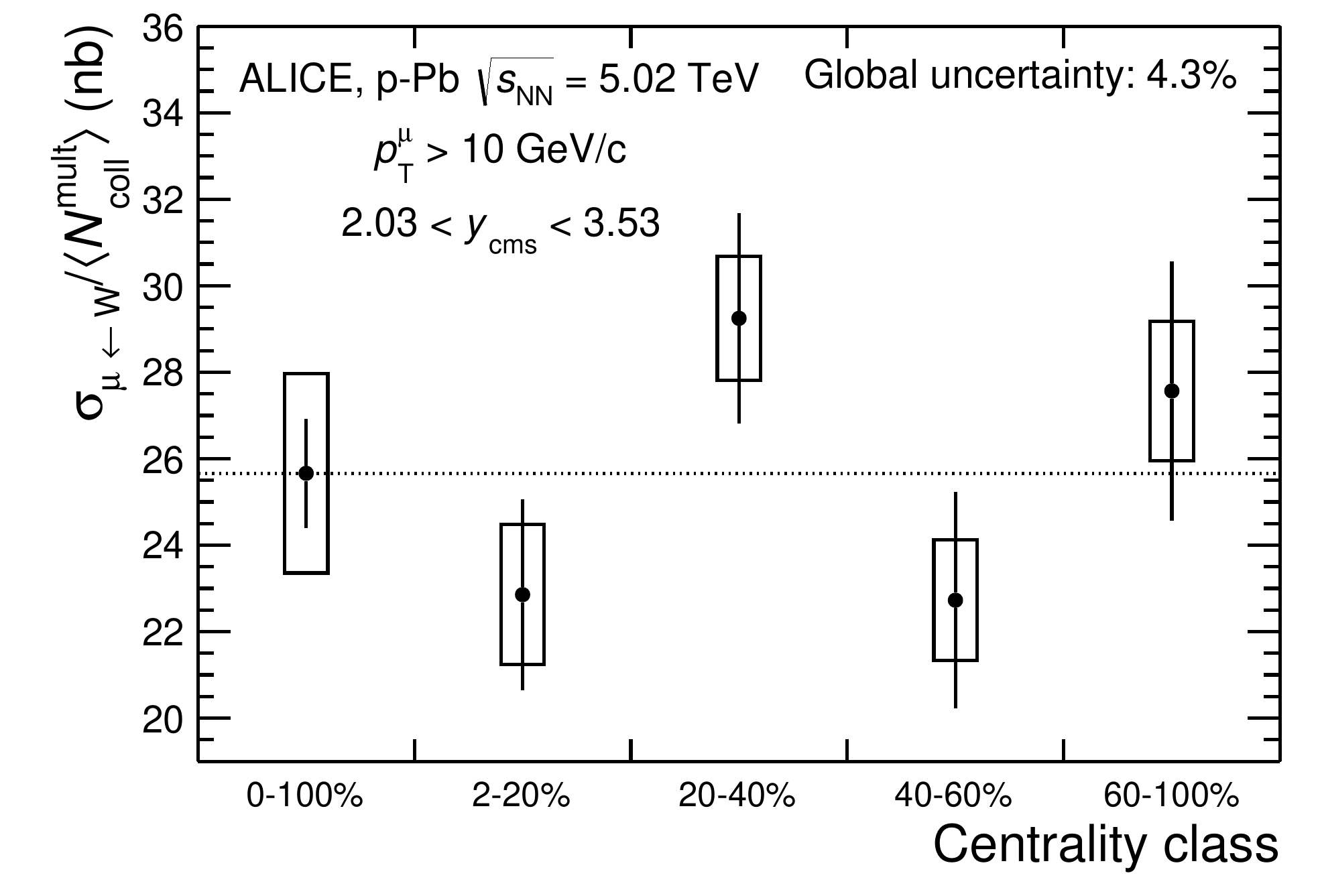}
\caption{Sum of the cross sections of positive and negative charge muons from W boson decays measured in \pPb collisions at \snn = 5.02~TeV in the rapidity region $-4.46 < \ycms < -2.96$ (left panel) and $2.03 < \ycms< 3.53$ (right panel) as a function of centrality. The cross sections are normalised by the number of binary collisions $\langle \ncmult \rangle$. The vertical bars (open boxes) represent the statistical (systematic) uncertainties. The correlated global uncertainties include the MB cross section, normalisation, \acceff corrections and tracking and trigger systematics. A dotted line is drawn at the value of the centrality-integrated cross section to guide the eye.}
\label{fig:wVsCentrality}
\end{figure}

The production of muons from W-boson decays with $\pt^\mu>10$\GeVc is studied as a function of the collision centrality.
Due to the limited statistics, the $\mu^+$ and $\mu^-$ results are summed together.
The resulting cross sections at backward and forward rapidities normalised by the average number of binary collisions~\cite{Adam:2014qja} are shown in the left and right panels of \fig{fig:wVsCentrality}, respectively.
The vertical bars represent the statistical uncertainties while the open boxes are the uncorrelated systematic ones.
The quadratic sum of the correlated systematic uncertainties on the MB cross section, normalisation, \acceff correction and tracking and trigger efficiency, which amounts to 4.8\% (4.3\%) in the Pb-going (p-going) sample, are quoted in the figure.

As discussed in the introduction, if the W boson production rate is consistent with geometric expectation, the production cross-section is expected to scale with the number of binary collisions for all centrality classes, provided that the centrality determination is not biased.
The measured centrality dependence is found to be compatible with a constant within uncertainties.

\section{Summary}\label{sec:conclusions}
The ALICE experiment has studied the W and Z-boson production at forward and backward rapidities in \pPb collisions at \snn = 5.02~TeV at the LHC.
The Z-boson cross section was measured in the dimuon decay channel with $\pt^\mu>20$\GeVc and $60<\invmass<120$\GeVcc.
The W-boson cross section and decay lepton charge asymmetry were measured in the muonic decay channel with $\pt^\mu>10$\GeVc.
The results are described by NLO pQCD calculations~\cite{Paukkunen:2010qg} as well as NNLO calculations using FEWZ~\cite{Gavin:2010az}, but the uncertainties on the measurement cannot constrain the nuclear modification of the PDFs.
W-boson production was also measured as a function of the event centrality, estimated from the energy deposited in the neutron zero degree calorimeters.
The cross section of muons from W-boson decays normalised by the number of binary nucleon-nucleon collisions is compatible with a constant within uncertainties.
Further measurements with better precision are needed to provide more stringent constraints on the nPDFs and on the binary scaling.

%% file: fa_2016-11-04.tex

The ALICE collaboration would like to thank Hannu Paukkunen for providing the pQCD calculations.
The ALICE Collaboration would like to thank all its engineers and technicians for their invaluable contributions to the construction of the experiment and the CERN accelerator teams for the outstanding performance of the LHC complex.
The ALICE Collaboration gratefully acknowledges the resources and support provided by all Grid centres and the Worldwide LHC Computing Grid (WLCG) collaboration.
The ALICE Collaboration acknowledges the following funding agencies for their support in building and running the ALICE detector:
A. I. Alikhanyan National Science Laboratory (Yerevan Physics Institute) Foundation (ANSL), State Committee of Science and World Federation of Scientists (WFS), Armenia;
Austrian Academy of Sciences and Nationalstiftung f\"{u}r Forschung, Technologie und Entwicklung, Austria;
Conselho Nacional de Desenvolvimento Cient\'{\i}fico e Tecnol\'{o}gico (CNPq), Universidade Federal do Rio Grande do Sul (UFRGS), Financiadora de Estudos e Projetos (Finep) and Funda\c{c}\~{a}o de Amparo \`{a} Pesquisa do Estado de S\~{a}o Paulo (FAPESP), Brazil;
Ministry of Science \& Technology of China (MSTC), National Natural Science Foundation of China (NSFC) and Ministry of Education of China (MOEC) , China;
Ministry of Science, Education and Sport and Croatian Science Foundation, Croatia;
Ministry of Education, Youth and Sports of the Czech Republic, Czech Republic;
The Danish Council for Independent Research | Natural Sciences, the Carlsberg Foundation and Danish National Research Foundation (DNRF), Denmark;
Helsinki Institute of Physics (HIP), Finland;
Commissariat \`{a} l'Energie Atomique (CEA) and Institut National de Physique Nucl\'{e}aire et de Physique des Particules (IN2P3) and Centre National de la Recherche Scientifique (CNRS), France;
Bundesministerium f\"{u}r Bildung, Wissenschaft, Forschung und Technologie (BMBF) and GSI Helmholtzzentrum f\"{u}r Schwerionenforschung GmbH, Germany;
Ministry of Education, Research and Religious Affairs, Greece;
National Research, Development and Innovation Office, Hungary;
Department of Atomic Energy Government of India (DAE), India;
Indonesian Institute of Science, Indonesia;
Centro Fermi - Museo Storico della Fisica e Centro Studi e Ricerche Enrico Fermi and Istituto Nazionale di Fisica Nucleare (INFN), Italy;
Institute for Innovative Science and Technology , Nagasaki Institute of Applied Science (IIST), Japan Society for the Promotion of Science (JSPS) KAKENHI and Japanese Ministry of Education, Culture, Sports, Science and Technology (MEXT), Japan;
Consejo Nacional de Ciencia (CONACYT) y Tecnolog\'{i}a, through Fondo de Cooperaci\'{o}n Internacional en Ciencia y Tecnolog\'{i}a (FONCICYT) and Direcci\'{o}n General de Asuntos del Personal Academico (DGAPA), Mexico;
Nationaal instituut voor subatomaire fysica (Nikhef), Netherlands;
The Research Council of Norway, Norway;
Commission on Science and Technology for Sustainable Development in the South (COMSATS), Pakistan;
Pontificia Universidad Cat\'{o}lica del Per\'{u}, Peru;
Ministry of Science and Higher Education and National Science Centre, Poland;
Korea Institute of Science and Technology Information and National Research Foundation of Korea (NRF), Republic of Korea;
Ministry of Education and Scientific Research, Institute of Atomic Physics and Romanian National Agency for Science, Technology and Innovation, Romania;
Joint Institute for Nuclear Research (JINR), Ministry of Education and Science of the Russian Federation and National Research Centre Kurchatov Institute, Russia;
Ministry of Education, Science, Research and Sport of the Slovak Republic, Slovakia;
National Research Foundation of South Africa, South Africa;
Centro de Aplicaciones Tecnol\'{o}gicas y Desarrollo Nuclear (CEADEN), Cubaenerg\'{\i}a, Cuba, Ministerio de Ciencia e Innovacion and Centro de Investigaciones Energ\'{e}ticas, Medioambientales y Tecnol\'{o}gicas (CIEMAT), Spain;
Swedish Research Council (VR) and Knut \& Alice Wallenberg Foundation (KAW), Sweden;
European Organization for Nuclear Research, Switzerland;
National Science and Technology Development Agency (NSDTA), Suranaree University of Technology (SUT) and Office of the Higher Education Commission under NRU project of Thailand, Thailand;
Turkish Atomic Energy Agency (TAEK), Turkey;
National Academy of  Sciences of Ukraine, Ukraine;
Science and Technology Facilities Council (STFC), United Kingdom;
National Science Foundation of the United States of America (NSF) and United States Department of Energy, Office of Nuclear Physics (DOE NP), United States of America.

%% file: Alice_Authorlist_2016-09-23_mod.tex


\begingroup
\small
\begin{flushleft}
J.~Adam$^\textrm{\scriptsize 39}$,
D.~Adamov\'{a}$^\textrm{\scriptsize 86}$,
M.M.~Aggarwal$^\textrm{\scriptsize 90}$,
G.~Aglieri Rinella$^\textrm{\scriptsize 35}$,
M.~Agnello$^\textrm{\scriptsize 113}$\textsuperscript{,}$^\textrm{\scriptsize 31}$,
N.~Agrawal$^\textrm{\scriptsize 48}$,
Z.~Ahammed$^\textrm{\scriptsize 137}$,
S.~Ahmad$^\textrm{\scriptsize 18}$,
S.U.~Ahn$^\textrm{\scriptsize 70}$,
S.~Aiola$^\textrm{\scriptsize 141}$
% ...
\renewcommand\labelenumi{\textsuperscript{\theenumi}~}

\section*{Affiliation notes}
\renewcommand\theenumi{\roman{enumi}}
\begin{Authlist}
\item \Adef{0}Deceased
\item \Adef{idp1808416}{Also at: Georgia State University, Atlanta, Georgia, United States}
\item \Adef{idp3223456}{Also at: Also at Department of Applied Physics, Aligarh Muslim University, Aligarh, India}
\item \Adef{idp3971968}{Also at: M.V. Lomonosov Moscow State University, D.V. Skobeltsyn Institute of Nuclear, Physics, Moscow, Russia}
\end{Authlist}

\section*{Collaboration Institutes}
\renewcommand\theenumi{\arabic{enumi}~}

$^{1}$A.I. Alikhanyan National Science Laboratory (Yerevan Physics Institute) Foundation, Yerevan, Armenia
\\
$^{2}$Benem\'{e}rita Universidad Aut\'{o}noma de Puebla, Puebla, Mexico
\\
$^{3}$Bogolyubov Institute for Theoretical Physics, Kiev, Ukraine
\\
$^{4}$Bose Institute, Department of Physics 
and Centre for Astroparticle Physics and Space Science (CAPSS), Kolkata, India
\\
$^{5}$Budker Institute for Nuclear Physics, Novosibirsk, Russia
\\
$^{6}$California Polytechnic State University, San Luis Obispo, California, United States
\\
$^{7}$Central China Normal University, Wuhan, China
\\
$^{8}$Centre de Calcul de l'IN2P3, Villeurbanne, Lyon, France
\\
$^{9}$Centro de Aplicaciones Tecnol\'{o}gicas y Desarrollo Nuclear (CEADEN), Havana, Cuba
\\
$^{10}$Centro de Investigaciones Energ\'{e}ticas Medioambientales y Tecnol\'{o}gicas (CIEMAT), Madrid, Spain
\\
$^{11}$Centro de Investigaci\'{o}n y de Estudios Avanzados (CINVESTAV), Mexico City and M\'{e}rida, Mexico
\\
$^{12}$Centro Fermi - Museo Storico della Fisica e Centro Studi e Ricerche ``Enrico Fermi', Rome, Italy
\\
$^{13}$Chicago State University, Chicago, Illinois, United States
\\
$^{14}$China Institute of Atomic Energy, Beijing, China
\\
$^{15}$Commissariat \`{a} l'Energie Atomique, IRFU, Saclay, France
\\
$^{16}$COMSATS Institute of Information Technology (CIIT), Islamabad, Pakistan
\\
$^{17}$Departamento de F\'{\i}sica de Part\'{\i}culas and IGFAE, Universidad de Santiago de Compostela, Santiago de Compostela, Spain
\\
$^{18}$Department of Physics, Aligarh Muslim University, Aligarh, India
\\
$^{19}$Department of Physics, Ohio State University, Columbus, Ohio, United States
\\
$^{20}$Department of Physics, Sejong University, Seoul, South Korea
\\
$^{21}$Department of Physics, University of Oslo, Oslo, Norway
\\
$^{22}$Department of Physics and Technology, University of Bergen, Bergen, Norway
\\
$^{23}$Dipartimento di Fisica dell'Universit\`{a} 'La Sapienza'
and Sezione INFN, Rome, Italy
\\
$^{24}$Dipartimento di Fisica dell'Universit\`{a}
and Sezione INFN, Cagliari, Italy
\\
$^{25}$Dipartimento di Fisica dell'Universit\`{a}
and Sezione INFN, Trieste, Italy
\\
$^{26}$Dipartimento di Fisica dell'Universit\`{a}
and Sezione INFN, Turin, Italy
\\
$^{27}$Dipartimento di Fisica e Astronomia dell'Universit\`{a}
and Sezione INFN, Bologna, Italy
\\
$^{28}$Dipartimento di Fisica e Astronomia dell'Universit\`{a}
and Sezione INFN, Catania, Italy
\\
$^{29}$Dipartimento di Fisica e Astronomia dell'Universit\`{a}
and Sezione INFN, Padova, Italy
\\
$^{30}$Dipartimento di Fisica `E.R.~Caianiello' dell'Universit\`{a}
and Gruppo Collegato INFN, Salerno, Italy
\\
$^{31}$Dipartimento DISAT del Politecnico and Sezione INFN, Turin, Italy
\\
$^{32}$Dipartimento di Scienze e Innovazione Tecnologica dell'Universit\`{a} del Piemonte Orientale and INFN Sezione di Torino, Alessandria, Italy
\\
$^{33}$Dipartimento Interateneo di Fisica `M.~Merlin'
and Sezione INFN, Bari, Italy
\\
$^{34}$Division of Experimental High Energy Physics, University of Lund, Lund, Sweden
\\
$^{35}$European Organization for Nuclear Research (CERN), Geneva, Switzerland
\\
$^{36}$Excellence Cluster Universe, Technische Universit\"{a}t M\"{u}nchen, Munich, Germany
\\
$^{37}$Faculty of Engineering, Bergen University College, Bergen, Norway
\\
$^{38}$Faculty of Mathematics, Physics and Informatics, Comenius University, Bratislava, Slovakia
\\
$^{39}$Faculty of Nuclear Sciences and Physical Engineering, Czech Technical University in Prague, Prague, Czech Republic
\\
$^{40}$Faculty of Science, P.J.~\v{S}af\'{a}rik University, Ko\v{s}ice, Slovakia
\\
$^{41}$Faculty of Technology, Buskerud and Vestfold University College, Tonsberg, Norway
\\
$^{42}$Frankfurt Institute for Advanced Studies, Johann Wolfgang Goethe-Universit\"{a}t Frankfurt, Frankfurt, Germany
\\
$^{43}$Gangneung-Wonju National University, Gangneung, South Korea
\\
$^{44}$Gauhati University, Department of Physics, Guwahati, India
\\
$^{45}$Helmholtz-Institut f\"{u}r Strahlen- und Kernphysik, Rheinische Friedrich-Wilhelms-Universit\"{a}t Bonn, Bonn, Germany
\\
$^{46}$Helsinki Institute of Physics (HIP), Helsinki, Finland
\\
$^{47}$Hiroshima University, Hiroshima, Japan
\\
$^{48}$Indian Institute of Technology Bombay (IIT), Mumbai, India
\\
$^{49}$Indian Institute of Technology Indore, Indore, India
\\
$^{50}$Indonesian Institute of Sciences, Jakarta, Indonesia
\\
$^{51}$Inha University, Incheon, South Korea
\\
$^{52}$Institut de Physique Nucl\'eaire d'Orsay (IPNO), Universit\'e Paris-Sud, CNRS-IN2P3, Orsay, France
\\
$^{53}$Institute for Nuclear Research, Academy of Sciences, Moscow, Russia
\\
$^{54}$Institute for Subatomic Physics of Utrecht University, Utrecht, Netherlands
\\
$^{55}$Institute for Theoretical and Experimental Physics, Moscow, Russia
\\
$^{56}$Institute of Experimental Physics, Slovak Academy of Sciences, Ko\v{s}ice, Slovakia
\\
$^{57}$Institute of Physics, Academy of Sciences of the Czech Republic, Prague, Czech Republic
\\
$^{58}$Institute of Physics, Bhubaneswar, India
\\
$^{59}$Institute of Space Science (ISS), Bucharest, Romania
\\
$^{60}$Institut f\"{u}r Informatik, Johann Wolfgang Goethe-Universit\"{a}t Frankfurt, Frankfurt, Germany
\\
$^{61}$Institut f\"{u}r Kernphysik, Johann Wolfgang Goethe-Universit\"{a}t Frankfurt, Frankfurt, Germany
\\
$^{62}$Institut f\"{u}r Kernphysik, Westf\"{a}lische Wilhelms-Universit\"{a}t M\"{u}nster, M\"{u}nster, Germany
\\
$^{63}$Instituto de Ciencias Nucleares, Universidad Nacional Aut\'{o}noma de M\'{e}xico, Mexico City, Mexico
\\
$^{64}$Instituto de F\'{i}sica, Universidade Federal do Rio Grande do Sul (UFRGS), Porto Alegre, Brazil
\\
$^{65}$Instituto de F\'{\i}sica, Universidad Nacional Aut\'{o}noma de M\'{e}xico, Mexico City, Mexico
\\
$^{66}$Institut Pluridisciplinaire Hubert Curien (IPHC), Universit\'{e} de Strasbourg, CNRS-IN2P3, Strasbourg, France
\\
$^{67}$iThemba LABS, National Research Foundation, Somerset West, South Africa
\\
$^{68}$Joint Institute for Nuclear Research (JINR), Dubna, Russia
\\
$^{69}$Konkuk University, Seoul, South Korea
\\
$^{70}$Korea Institute of Science and Technology Information, Daejeon, South Korea
\\
$^{71}$KTO Karatay University, Konya, Turkey
\\
$^{72}$Laboratoire de Physique Corpusculaire (LPC), Clermont Universit\'{e}, Universit\'{e} Blaise Pascal, CNRS--IN2P3, Clermont-Ferrand, France
\\
$^{73}$Laboratoire de Physique Subatomique et de Cosmologie, Universit\'{e} Grenoble-Alpes, CNRS-IN2P3, Grenoble, France
\\
$^{74}$Laboratori Nazionali di Frascati, INFN, Frascati, Italy
\\
$^{75}$Laboratori Nazionali di Legnaro, INFN, Legnaro, Italy
\\
$^{76}$Lawrence Berkeley National Laboratory, Berkeley, California, United States
\\
$^{77}$Moscow Engineering Physics Institute, Moscow, Russia
\\
$^{78}$Nagasaki Institute of Applied Science, Nagasaki, Japan
\\
$^{79}$National Centre for Nuclear Studies, Warsaw, Poland
\\
$^{80}$National Institute for Physics and Nuclear Engineering, Bucharest, Romania
\\
$^{81}$National Institute of Science Education and Research, Bhubaneswar, India
\\
$^{82}$National Research Centre Kurchatov Institute, Moscow, Russia
\\
$^{83}$Niels Bohr Institute, University of Copenhagen, Copenhagen, Denmark
\\
$^{84}$Nikhef, Nationaal instituut voor subatomaire fysica, Amsterdam, Netherlands
\\
$^{85}$Nuclear Physics Group, STFC Daresbury Laboratory, Daresbury, United Kingdom
\\
$^{86}$Nuclear Physics Institute, Academy of Sciences of the Czech Republic, \v{R}e\v{z} u Prahy, Czech Republic
\\
$^{87}$Oak Ridge National Laboratory, Oak Ridge, Tennessee, United States
\\
$^{88}$Petersburg Nuclear Physics Institute, Gatchina, Russia
\\
$^{89}$Physics Department, Creighton University, Omaha, Nebraska, United States
\\
$^{90}$Physics Department, Panjab University, Chandigarh, India
\\
$^{91}$Physics Department, University of Athens, Athens, Greece
\\
$^{92}$Physics Department, University of Cape Town, Cape Town, South Africa
\\
$^{93}$Physics Department, University of Jammu, Jammu, India
\\
$^{94}$Physics Department, University of Rajasthan, Jaipur, India
\\
$^{95}$Physikalisches Institut, Eberhard Karls Universit\"{a}t T\"{u}bingen, T\"{u}bingen, Germany
\\
$^{96}$Physikalisches Institut, Ruprecht-Karls-Universit\"{a}t Heidelberg, Heidelberg, Germany
\\
$^{97}$Physik Department, Technische Universit\"{a}t M\"{u}nchen, Munich, Germany
\\
$^{98}$Purdue University, West Lafayette, Indiana, United States
\\
$^{99}$Pusan National University, Pusan, South Korea
\\
$^{100}$Research Division and ExtreMe Matter Institute EMMI, GSI Helmholtzzentrum f\"ur Schwerionenforschung, Darmstadt, Germany
\\
$^{101}$Rudjer Bo\v{s}kovi\'{c} Institute, Zagreb, Croatia
\\
$^{102}$Russian Federal Nuclear Center (VNIIEF), Sarov, Russia
\\
$^{103}$Saha Institute of Nuclear Physics, Kolkata, India
\\
$^{104}$School of Physics and Astronomy, University of Birmingham, Birmingham, United Kingdom
\\
$^{105}$Secci\'{o}n F\'{\i}sica, Departamento de Ciencias, Pontificia Universidad Cat\'{o}lica del Per\'{u}, Lima, Peru
\\
$^{106}$Sezione INFN, Bari, Italy
\\
$^{107}$Sezione INFN, Bologna, Italy
\\
$^{108}$Sezione INFN, Cagliari, Italy
\\
$^{109}$Sezione INFN, Catania, Italy
\\
$^{110}$Sezione INFN, Padova, Italy
\\
$^{111}$Sezione INFN, Rome, Italy
\\
$^{112}$Sezione INFN, Trieste, Italy
\\
$^{113}$Sezione INFN, Turin, Italy
\\
$^{114}$SSC IHEP of NRC Kurchatov institute, Protvino, Russia
\\
$^{115}$Stefan Meyer Institut f\"{u}r Subatomare Physik (SMI), Vienna, Austria
\\
$^{116}$SUBATECH, Ecole des Mines de Nantes, Universit\'{e} de Nantes, CNRS-IN2P3, Nantes, France
\\
$^{117}$Suranaree University of Technology, Nakhon Ratchasima, Thailand
\\
$^{118}$Technical University of Ko\v{s}ice, Ko\v{s}ice, Slovakia
\\
$^{119}$Technical University of Split FESB, Split, Croatia
\\
$^{120}$The Henryk Niewodniczanski Institute of Nuclear Physics, Polish Academy of Sciences, Cracow, Poland
\\
$^{121}$The University of Texas at Austin, Physics Department, Austin, Texas, United States
\\
$^{122}$Universidad Aut\'{o}noma de Sinaloa, Culiac\'{a}n, Mexico
\\
$^{123}$Universidade de S\~{a}o Paulo (USP), S\~{a}o Paulo, Brazil
\\
$^{124}$Universidade Estadual de Campinas (UNICAMP), Campinas, Brazil
\\
$^{125}$Universidade Federal do ABC, Santo Andre, Brazil
\\
$^{126}$University of Houston, Houston, Texas, United States
\\
$^{127}$University of Jyv\"{a}skyl\"{a}, Jyv\"{a}skyl\"{a}, Finland
\\
$^{128}$University of Liverpool, Liverpool, United Kingdom
\\
$^{129}$University of Tennessee, Knoxville, Tennessee, United States
\\
$^{130}$University of the Witwatersrand, Johannesburg, South Africa
\\
$^{131}$University of Tokyo, Tokyo, Japan
\\
$^{132}$University of Tsukuba, Tsukuba, Japan
\\
$^{133}$University of Zagreb, Zagreb, Croatia
\\
$^{134}$Universit\'{e} de Lyon, Universit\'{e} Lyon 1, CNRS/IN2P3, IPN-Lyon, Villeurbanne, Lyon, France
\\
$^{135}$Universit\`{a} di Brescia, Brescia, Italy
\\
$^{136}$V.~Fock Institute for Physics, St. Petersburg State University, St. Petersburg, Russia
\\
$^{137}$Variable Energy Cyclotron Centre, Kolkata, India
\\
$^{138}$Warsaw University of Technology, Warsaw, Poland
\\
$^{139}$Wayne State University, Detroit, Michigan, United States
\\
$^{140}$Wigner Research Centre for Physics, Hungarian Academy of Sciences, Budapest, Hungary
\\
$^{141}$Yale University, New Haven, Connecticut, United States
\\
$^{142}$Yonsei University, Seoul, South Korea
\\
$^{143}$Zentrum f\"{u}r Technologietransfer und Telekommunikation (ZTT), Fachhochschule Worms, Worms, Germany
\endgroup